\documentclass[a4paper,11pt]{article}
\pdfoutput=1

\usepackage{jheparxiv} 
\usepackage[T1]{fontenc}
\usepackage[utf8]{inputenc}
\usepackage{graphicx}
\usepackage{amsmath}
\usepackage{amssymb}
\usepackage{dsfont}
\usepackage{latexsym}
\usepackage{mathrsfs}
\usepackage{braket}		
\usepackage{hyperref}
\usepackage{xcolor}
\usepackage{twistor}
\usepackage{mathtools}
\usepackage{tikz-cd}

\newcommand{\msf}[1]{\mathsf{#1}}

\newcommand{\eps}{\epsilon}
\newcommand{\veps}{\varepsilon}

\newcommand{\sZ}{\mathsf{Z}}
\newcommand{\sW}{\mathsf{W}}

\newcommand{\bL}{\mathbf{L}}

\newcommand{\defeq}{\vcentcolon=}
\newcommand{\sa}{\msf{a}}
\newcommand{\bsa}{\bar{\msf{a}}}

\title{Ambidextrous light transforms for celestial amplitudes}
\author[a]{Atul Sharma}

\affiliation[a]{The Mathematical Institute,\\ University of Oxford, Woodstock Road, OX2 6GG, United Kingdom}

\emailAdd{atul.sharma@maths.ox.ac.uk}

\abstract{Low multiplicity celestial amplitudes of gluons and gravitons tend to be distributional in the celestial coordinates $z,\bar z$. We provide a new systematic remedy to this situation by studying celestial amplitudes in a basis of light transformed boost eigenstates. Motivated by a novel equivalence between light transforms and Witten's half-Fourier transforms to twistor space, we light transform every positive helicity state in the coordinate $z$ and every negative helicity state in $\bar z$. With examples, we show that this ``ambidextrous'' prescription beautifully recasts two- and three-point celestial amplitudes in terms of standard conformally covariant structures. These are used to extract examples of celestial OPE for light transformed operators. We also study such amplitudes at higher multiplicity by constructing the Grassmannian representation of tree-level gluon celestial amplitudes as well as their light transforms. The formulae for $n$-point N$^{k-2}$MHV amplitudes take the form of Euler-type integrals over regions in $\Gr(k,n)$ cut out by positive energy constraints.}

\begin{document}

\maketitle

\section{Introduction}

What are the fundamental degrees of freedom of flat space physics? In celestial holography, this question takes the guise of a search for the spectrum of local operators in celestial conformal field theories (CCFT). Knowing the local operators and their OPE algebra, one could hope to provide a completely holographic description of scattering amplitudes in flat space. But even the definition of a local operator gets muddied by the existence of integral transforms like the light and shadow transforms. These map local operators to non-local operators that superficially appear to be locally defined \cite{Simmons-Duffin:2012juh,Kravchuk:2018htv}.

Until recently, the working hypothesis has been that every boost eigenstate of a particle in the bulk is dual to a local operator living in the CCFT \cite{Pasterski:2016qvg, Pasterski:2017kqt}. However, the CCFT stress tensor is naturally defined to be the shadow transform of a conformally soft $\Delta=0$ graviton \cite{Kapec:2016jld, Donnay:2018neh, Adamo:2019ipt,Puhm:2019zbl, Fotopoulos:2019vac}. This hints at the possibility that at least soft shadow modes may be dual to actual local operators. Recent work \cite{Crawley:2021ivb} has further exemplified the significance of the shadow transformed operators for defining a meaningful inner product and 2-point functions on the CCFT Hilbert space. More generally, they can be used to convert the distribution-valued low multiplicity celestial amplitudes to the standard conformally covariant structures familiar from 2d CFT. For instance, this idea has been applied to derive conformal block expansions for gluon amplitudes \cite{Fan:2021isc}. The shadow operators also automatically occur in the conformal multiplets associated to soft modes \cite{Pasterski:2021fjn, Pasterski:2021dqe}.

In this work, we wish to focus on a refinement of the shadow transform in Lorentzian signature, namely the light transform. The authors of \cite{Atanasov:2021oyu} studied CCFT on the celestial torus of flat space $\R^{2,2}$ with a split signature metric. This is a Lorentzian CFT and hence contains light ray operators with analytically continued spins \cite{Kravchuk:2018htv}. The simplest of these arise from light transforming local operators along one of the two null directions on the celestial torus. They were front and center in the conformal block expansion of 4-scalar celestial amplitudes \cite{Atanasov:2021cje}, and also played a role in the recent reorganization of asymptotic symmetries of quantum gravity \cite{Strominger:2021lvk}. Hence, even by themselves these insights provide enough tantalizing motivation for us to explore celestial amplitudes in a basis of light transformed boost eigenstates. 

At this stage, a natural conundrum rears its head. \emph{Which states should we light transform and which null geodesics do we integrate along?} In the case of the shadow transform, previous studies like \cite{Fan:2021isc,Crawley:2021ivb} have tried out various physically or computationally motivated choices of which states to transform. But a general prescription is still lacking. In order to remedy the analogous situation for light transforms, we look to gain further intuition about them. Once again, a useful place to start is momentum space. One could ask the even more basic question: \emph{what is the analogue of light transforms for momentum eigenstates?} It turns out that answering this question thrusts us head first into the world of half-Fourier transforms and twistor space!

In the following, we will describe how light transformed states automatically make an appearance in the context of relatively well-studied notions like twistor eigenstates and amplitudes in twistor space \cite{Adamo:2017qyl,Mason:2009sa,Adamo:2011pv}. Over the years, twistors have played a starring role in the development of the modern geometric underpinnings of scattering theory. Staying in split signature, Witten's half-Fourier transform originally acted on massless momentum eigenstates $|\lambda,\bar\lambda\ra$ --- labeled by spinor-helicity variables $\lambda_\al,\bar\lambda_{\dal}$ --- and Fourier transformed $\bar\lambda_{\dal}$ to a twistor coordinate $\mu^{\dal}$. The observation that amplitudes of the resulting ``twistor eigenstates'' $|\lambda,\mu\ra$ localize on rational curves in twistor space was one of the critical motivations for twistor strings \cite{Witten:2003nn}. It was also the progenitor of various link representations and Grassmannian formulae \cite{ArkaniHamed:2009si,ArkaniHamed:2009dn, He:2012er, Cachazo:2012pz, Arkani-Hamed:2012zlh}.

We will generalize this half-Fourier transform to act on boost eigenstates. For this purpose, we first define a ``half-Mellin'' transform. For example, we set $\lambda_\al = \pm t\,(z,1)$ in affine coordinates and Mellin transform over $t\in\R_+ \equiv (0,\infty)$ while leaving $\bar\lambda_{\dal}$ untouched. This produces a mild generalization of the standard boost eigenstate that now depends on generic $\bar\lambda_{\dal}$. Following this, we can freely Fourier transform in the $\bar\lambda_{\dal}$ to define ``conformal primary twistor eigenstates'' labeled by $z$ and $\mu^{\dal}$. Celestial amplitudes can be recovered from the amplitudes of such primary twistor eigenstates via inverse half-Fourier transforms and the substitutions $\bar\lambda_{\dal} = \pm(\bar z,1)$. Whereas twistor amplitudes can be found by inverting the half-Mellin transforms. More encouragingly, we will show that these primary twistor eigenstates nicely decompose into a two-term linear combination \eqref{momtoZ4} of positive and negative frequency light transformed boost eigenstates. Thus the light transformed states could even be thought of as being more fundamental, with the twistor eigenstates and amplitudes being obtainable through appropriate finite linear combinations.

This raises the possibility that celestial holography could be the ultimate origin of the observed simplicity of amplitudes in twistor space. Along these lines, we propose to study celestial amplitudes in an ambidextrous basis of light transformed states (instead of twistor eigenstates). In \cite{ArkaniHamed:2009si}, the authors half-Fourier transformed momentum space amplitudes in $\bar\lambda_{\dal}$ for all the negative helicity particles and in $\lambda_\al$ for the positive helicity ones. This led to an ambidextrous presentation of twistor amplitudes that was shown to be derivable from a beautiful Grassmannian integral formula called the link representation. Taking inspiration from this, we experiment with this philosophy by applying it to celestial amplitudes. But instead of half-Fourier transforming, we \emph{light transform} the negative helicity modes in $\bar z$ and the positive helicity modes in $z$. 

The main happy consequence of this ambidextrous choice of transforms is that the light ray integrals soak up all the residual momentum conserving delta functions that are ubiquitous in low multiplicity celestial amplitudes \cite{Pasterski:2017ylz, Puhm:2019zbl, Schreiber:2017jsr, Casali:2020uvr}, leading to CFT correlators of the more recognizable kind. In particular, this helps demystify the celestial OPE limit. When computing the OPE limit of gluon and graviton celestial amplitudes, one often simply Mellin transforms the collinear expansion of momentum space amplitudes \cite{Fan:2019emx, Pate:2019lpp}. This leads to OPE coefficients given by Euler Beta functions. But it is never clear how these Beta functions could emerge directly from the distributional expressions for the low multiplicity celestial amplitudes. With some easy examples, we show that 3-point light transformed celestial amplitudes manifestly contain factors of such OPE coefficients and cleanly reduce to the corresponding 2-point amplitudes in the OPE limit. This allows us to extract the leading celestial OPE of light transformed gluons and gravitons without appealing to their collinear behavior in momentum space. 

We then finish by initiating a study of Grassmannian formulae for gluon celestial amplitudes as well as their light transformed cousins. The $n$-gluon momentum space link representation contains $2n$ delta functions. Celestial amplitudes are found by localizing the Mellin integrals in their definition against $n$ of these delta functions. The resulting formula \eqref{cAnk1} contains $n$ leftover delta functions and interpolates between ``integrals that can be localized'' and integrals of Euler-type. On performing the proposed $n$ light transforms against the rest of these delta functions, we discover a link representation \eqref{Lnk} that contains purely Euler-type integrals over $\Gr(k,n)$ at MHV degree $k-2$. We have not performed a detailed study of their contours of integration inside the complex Grassmannian (beyond what BCFW recursion already tells us), but the resulting formulae do contain step functions imposing ``positive energy'' conditions. Similar step functions are encountered in all other studies of celestial amplitudes (see for instance \cite{Schreiber:2017jsr}), but in our case they also involve the Grassmannian integration variables and constrain the integration contours.

To summarize, we begin with the necessary background about celestial amplitudes in section \ref{sec:pre}. In section \ref{sec:lt}, we recall the definitions of light transforms on the celestial torus and illustrate them by computing light transformed conformal primary wavefunctions for scalars, gluons and gravitons. In section \ref{sec:twistor}, we describe half-Mellin and half-Fourier transforms that take boost eigenstates to the appropriate notion of conformal primary twistor eigenstates. We also show that these twistor eigenstates are equivalent to a linear combination of two light transformed boost eigenstates. Motivated by these ideas, in section \ref{sec:ltamp} we study 2- and 3-point examples of light transformed celestial amplitudes in ambidextrous bases of gluons and gravitons. This also allows for a basic study of the celestial OPE of light transformed operators. In section \ref{sec:grass} we construct the all-multiplicity Grassmannian formulae for gluon celestial amplitudes and their light transforms, concluding in section \ref{discussion} with a look toward the future.


\section{Preliminaries}
\label{sec:pre}

Boost eigenstates are defined as Mellin transforms of momentum eigenstates \cite{Pasterski:2017kqt}. To set up the correspondence, let $|\lambda,\bar\lambda,\ell\ra$ denote the momentum eigenstate of a massless particle with momentum $p_{\al\dal} = \lambda_\al\bar\lambda_{\dal}$ and helicity $\ell$. The $2$-spinors $\lambda_\al$, $\bar\lambda_{\dal}$ denote its spinor-helicity variables. Working in Klein space $\R^{2,2}$ \cite{Atanasov:2021oyu} with flat metric of split signature $(+\,+\,-\;-)$, we take them to be real-valued and independent of each other. Under a little group scaling, $|\lambda,\bar\lambda,\ell\ra$ transforms as
\begin{equation}\label{lg}
    |r\,\lambda,r^{-1}\,\bar\lambda,\ell\ra = r^{-2\ell}\,|\lambda,\bar\lambda,\ell\ra\,,\qquad r\in\R^*\,.
\end{equation}
Next, without loss of generality, we can decompose
\begin{equation}\label{shvar}
    \lambda_\al = t\,\zeta_\al \equiv t\begin{pmatrix}z\\1\end{pmatrix}\,,\qquad\bar\lambda_{\dal} = \bar t\,\bar\zeta_{\dal} \equiv \bar t \begin{pmatrix}\bar z\\1\end{pmatrix}\,,
\end{equation}
where $z,\bar z\in\R$ are the celestial coordinates and $t,\bar t\in\R^*$ act as overall scalings. We can also view $\zeta_\al,\bar\zeta_{\dal}$ as homogeneous coordinates on the celestial torus $\RP^1\times\RP^1$.

We define a boost eigenstate of boost weight $\Delta$ and spin weight $\ell$ by a Mellin integral
\begin{equation}\label{cpstate}
    |\zeta,\bar\zeta,h,\bar h,\eps\ra \vcentcolon = \int_{\R_+}\frac{\d\omega}{\omega}\,\omega^\Delta\;|\sqrt{\omega}\,\zeta,\eps\sqrt{\omega}\,\bar\zeta,\ell\ra\,,
\end{equation}
having labeled it by its conformal weights $(h,\bar h) = (\frac{\Delta+\ell}{2},\frac{\Delta-\ell}{2})$. We also write this as $|z,\bar z,h,\bar h,\eps\ra$ when working with affine coordinates $z,\bar z$. The sign $\eps\in\{\pm1\}$ indicates whether the displayed momentum eigenstate on the right is a positive or negative frequency state. $|\zeta,\bar\zeta,h,\bar h,\eps\ra$ transforms covariantly under celestial conformal rescalings:
\begin{equation}
    |s\,\zeta,\bar s\,\bar\zeta,h,\bar h,\eps\ra = s^{-2h}\,\bar s^{-2\bar h}\,|\zeta,\bar\zeta,h,\bar h,\eps\ra\,,\qquad s,\bar s\in\R_+\,.
\end{equation}
The CCFT operator dual to $|\zeta,\bar\zeta,h,\bar h,\eps\ra$ will be called $\cO_{h,\bar h}^{\eps}(\zeta,\bar\zeta)$ or equivalently $\cO_{h,\bar h}^{\eps}(z,\bar z)$. It can be viewed as a field of homogeneity $(-2h,-2\bar h)$ in $(\zeta_\al,\bar\zeta_{\dal})$ in the embedding formalism for CFTs \cite{Costa:2011mg}.

Wherever convenient, we will work in this embedding space formalism. That is, instead of working with affine coordinates $z,\bar z$, we continue working with homogeneous coordinates $\zeta_\al,\bar\zeta_{\dal}$. Conformal transformations act as linear $\SL(2,\R)$ transformations; for instance,
\begin{equation}
    \zeta_\al = \begin{pmatrix}z\\1\end{pmatrix}\mapsto\begin{pmatrix}a&&b\\c&&d\end{pmatrix}\begin{pmatrix}z\\1\end{pmatrix} = (cz+d)\begin{pmatrix}\frac{az+b}{cz+d}\\1\end{pmatrix}\,,\; \text{etc.}
\end{equation}
And conformally covariant functions are always functions of the spinor contractions
\begin{equation}
    \la\zeta\,\zeta'\ra \equiv \zeta^\al\,\zeta'_\al = z'-z\,,\qquad[\bar\zeta\,\bar\zeta'] \equiv \bar\zeta^{\dal}\,\bar\zeta'_{\dal} = \bar z'-\bar z\,,
\end{equation}
having set $\zeta'_\al = (z',1)$ and $\bar\zeta'_{\dal} = (\bar z',1)$. Spinor indices are raised with the conventions $\zeta^\al = \eps^{\al\beta}\zeta_\beta$, $\bar\zeta^{\dal} = \eps^{\dal\dot\beta}\bar\zeta_{\dot\beta}$ with $\eps^{\al\beta}$, $\eps^{\dal\dot\beta}$ being Levi-Civita symbols.

Celestial amplitudes $\cA_n$ of massless particles are defined as the scattering amplitudes of such states. They are obtained by Mellin transforming $n$-point momentum space amplitudes $A_n(\lambda_i,\bar\lambda_i,\ell_i)$ \cite{Pasterski:2017kqt,Pasterski:2017ylz}:
\begin{equation}\label{celamp}
    \cA_n(\zeta_i,\bar\zeta_i,\Delta_i,\ell_i,\eps_i) = \int_{\R_+^n}\prod_{j=1}^n\frac{\d\omega_j}{\omega_j}\,\omega_j^{\Delta_j}\;A_n(\sqrt{\omega_i}\,\zeta_i,\eps_i\,\sqrt{\omega_i}\,\bar\zeta_i,\ell_i)\,.
\end{equation}
Here, $A_n$ includes the momentum conserving delta functions. At 5 and higher points, one can completely localize these delta functions against the Mellin integrals. But at 4 points or less, the resulting amplitudes are generically distributional in the $z_i,\bar z_i$ \cite{Pasterski:2017ylz}.


\section{Light transform}
\label{sec:lt}

In this section, we set up notation and review the definitions of light transforms along light rays on the celestial torus $\RP^1\times\RP^1$. With a Lorentzian celestial metric $\d s^2 = \d z\,\d\bar z$ there are only two null geodesics --- one spanning each copy of $\RP^1$ --- so there are only two possibilities. To gain familiarity with this construction, we compute the light transforms of scalar, gluon and graviton conformal primary wavefunctions. 

\medskip

The light ray transforms of CCFT operators $\cO_{h,\bar h}^{\eps}$ are defined as follows \cite{Kravchuk:2018htv,Atanasov:2021cje}:\footnote{Color indices on gluon operators will be reinstated as and when needed in what follows.}
\begin{align}
    \bL[\cO_{h,\bar h}^{\eps}](\zeta,\bar\zeta) &\vcentcolon = \int_{\RP^1}\frac{\D\zeta'}{\la\zeta\,\zeta'\ra^{2-2h}}\,\cO_{h,\bar h}^{\eps}(\zeta',\bar\zeta) \equiv \int_{\R}\frac{\d z'}{(z'-z)^{2-2h}}\,\cO_{h,\bar h}^{\eps}(z',\bar z)\,,\\
    \bar\bL[\cO_{h,\bar h}^{\eps}](\zeta,\bar\zeta) &\vcentcolon = \int_{\RP^1}\frac{\D\bar\zeta'}{[\bar\zeta\,\bar\zeta']^{2-2\bar h}}\,\cO_{h,\bar h}^{\eps}(\zeta,\bar\zeta')\equiv \int_{\R}\frac{\d\bar z'}{(\bar z'-\bar z)^{2-2\bar h}}\,\cO_{h,\bar h}^{\eps}(z,\bar z')\,.
\end{align}
The canonical integration measures on $\RP^1$ are
\begin{equation}
    \D\zeta' \equiv \la\zeta'\,\d\zeta'\ra = \d z'\,,\qquad\D\bar\zeta' \equiv [\bar\zeta'\,\d\bar\zeta'] = \d\bar z'\,.
\end{equation}
Here we have also expressed the measures in standard affine coordinates $\zeta'_\al = (z',1)$, etc., but we will later see that it will be much more convenient to directly choose conformal frames on $\RP^1$ that maximally simplify the integrands. Using the homogeneity of $\cO_{h,\bar h}^{\eps}$ in $\zeta'$, $\bar\zeta'$, it is easily verified that the integrands are invariant under $\GL(1,\R)$ rescalings. Thus the projective integrals over $\RP^1$ are well-defined and one can freely choose any useful conformal frame.

The states dual to these light transforms can be constructed in tandem:
\begin{align}
    |\zeta,\bar\zeta,1-h,\bar h,\eps\ra_\bL &\vcentcolon = \int_{\RP^1}\frac{\D\zeta'}{\la\zeta\,\zeta'\ra^{2-2h}}\,|\zeta',\bar\zeta,h,\bar h,\eps\ra \equiv \int_{\R}\frac{\d z'}{(z'-z)^{2-2h}}\,|z',\bar z,h,\bar h,\eps\ra\,,\label{Lstate}\\
    |\zeta,\bar\zeta,h,1-\bar h,\eps\ra_{\bar\bL} &\vcentcolon = \int_{\RP^1}\frac{\D\bar\zeta'}{[\bar\zeta\,\bar\zeta']^{2-2\bar h}}\,|\zeta,\bar\zeta',h,\bar h,\eps\ra \equiv \int_{\R}\frac{\d\bar z'}{(\bar z'-\bar z)^{2-2\bar h}}\,|z,\bar z',h,\bar h,\eps\ra\,.\label{Lbstate}
\end{align}
These states are conformal primary with weights $(1-h,\bar h)$ and $(h,1-\bar h)$ respectively. To distinguish them from the original boost eigenstates $|\zeta,\bar\zeta,1-h,\bar h,\eps\ra$, $|\zeta,\bar\zeta,h,1-\bar h,\eps\ra$, we have tagged them with subscripts $\bL$ or $\bar\bL$ as appropriate. Note also their quintessential property of having complex spin $\pm(1-\Delta)$ when $\Delta$ lies on the principle series $1+\im\,\R$.


\subsection{Scalars}

The conformal primary spin 0 wavefunction is given by \cite{Pasterski:2017kqt}
\begin{equation}
    \phi_{\Delta}^\eps(x\,|\,\zeta,\bar\zeta) = \int_{\R_+}\frac{\d\omega}{\omega}\,\omega^\Delta\,\e^{-\im\eps\omega x^{\al\dal}\zeta_\al\bar\zeta_{\dal}-\veps\omega} = \frac{(\im\,\eps)^{-\Delta}\,\Gamma(\Delta)}{(\la\zeta|x|\bar\zeta]-\im\,\eps\,\veps)^\Delta}\,,
\end{equation}
where we have abbreviated $x^{\al\dal}\zeta_\al\bar\zeta_{\dal}\equiv\la\zeta|x|\bar\zeta]$. The $\im\,\veps$-prescription here is required purely to make the Mellin integral converge. We will drop it in most of what follows. The light transform $\bL$ of this scalar wavefunction was computed in affine coordinates in \cite{Atanasov:2021cje}, but it is nonetheless instructive to do such computations in spinor notation. We explicitly calculate the transform $\bar\bL$.

We start with the light ray integral
\begin{equation}
    \bar\bL[\phi_{\Delta}^\eps](x\,|\,\zeta,\bar\zeta)=\int_{\RP^1}\frac{\D\bar\zeta'}{[\bar\zeta\,\bar\zeta']^{2-\Delta}}\,\frac{(\im\,\eps)^{-\Delta}\,\Gamma(\Delta)}{\la\zeta|x|\bar\zeta']^\Delta}\,.
\end{equation}
This integral is conformally covariant but only depends on two fixed points $\bar\zeta_{\dal}$ and $x^\al{}_{\dal}\zeta_\al$ on $\RP^1$. Thus, it has to be proportional to some power of $1/\la\zeta|x|\bar\zeta]$.\footnote{This is similar to how a conformally covariant 2-point function depending on two points $z_1$, $z_2$ has to be a function of $z_1-z_2$. Alternatively it could be a conformally covariant distribution.} However, unless $\Delta=1$, it also has different scaling weights in $\zeta_\al$ and $\bar\zeta_{\dal}$, which is impossible for something that is only a function of $\la\zeta|x|\bar\zeta]$.  Naively, this is only possible if the constant of proportionality was zero or divergent. To be general, we will work with the latter case.
Following \cite{Atanasov:2021cje}, one tames this integral using a regulator that partially breaks conformal covariance. Consider the regulated definition
\begin{equation}\label{Lbarsc}
    \bar\bL[\phi_{\Delta}^\eps](x\,|\,\zeta,\bar\zeta)=\lim_{\delta\to0}\int_{\RP^1}\frac{\D\bar\zeta'}{[\bar\zeta\,\bar\zeta']^{2-\Delta-\delta}}\,\frac{(\im\,\eps)^{-\Delta}\,\Gamma(\Delta)}{\la\zeta|x|\bar\zeta']^\Delta}\,\frac{1}{[\bar\iota\,\bar\zeta']^{\delta}}\,.
\end{equation}
In writing this, we have made the choice of a third ``reference point'' $\bar\iota_{\dal}$ on $\RP^1$. The integrand has been regulated in a way so as to preserve its invariance under $\GL(1,\R)$ scalings $\bar\zeta'_{\dal}\mapsto r\,\bar\zeta'_{\dal}$. As the exponent $\delta\to0$ --- or alternatively $\bar\iota_{\dal}\to\bar\zeta_{\dal}$ --- we get back our original integral.
We will be able to take the limit $\delta\to0$ after evaluating the integral.

To evaluate \eqref{Lbarsc}, we choose a conformal frame in which the three points $\bar\zeta_{\dal}$, $x^\al{}_{\dal}\zeta_\al$ and $\bar\iota_{\dal}$ on $\RP^1$ are respectively mapped to $0$, $1$ and $\infty$:
\begin{equation}\label{zetabarsub}
    \bar\zeta'_{\dal} = \bar\zeta_{\dal} - \bar z'\,\frac{\la\zeta|x|\bar\zeta]}{\la\zeta|x|\bar\iota]}\,\bar\iota_{\dal}\,.
\end{equation}
With this integral substitution, we are left with an affine integral over the parameter $\bar z'$,
\begin{multline}
    \lim_{\delta\to0}\frac{(\im\,\eps)^{-\Delta}\,\Gamma(\Delta)}{\la\zeta|x|\bar\zeta]^{1-\delta}}\,\frac{\la\zeta|x|\bar\iota]^{1-\Delta-\delta}}{[\bar\iota\,\bar\zeta]^{1-\Delta}}\int_\R\d\bar z'\;\bar z'^{\Delta-2+\delta}\,(1-\bar z')^{-\Delta}\\
    = \frac{(\im\,\eps)^{-\Delta}\,\Gamma(\Delta)}{\la\zeta|x|\bar\zeta]}\left(\frac{\la\zeta|x|\bar\iota]}{[\bar\iota\,\bar\zeta]}\right)^{1-\Delta}\lim_{\delta\to0}\frac{-2\pi\im\,\Gamma(1-\delta)}{\Gamma(\Delta)\,\Gamma(2-\Delta)}\,.
\end{multline}
The $\bar z'$ integral has been performed by breaking it up into the three ranges $(-\infty,0)$, $(0,1)$ and $(1,\infty)$ and using the Euler integral representation of the Beta function. As we will use it multiple times below, we also note the more general integration identity,\footnote{This is obtained for Re$(a)<1$, Re$(b)<1$ and Re$(a+b)>1$, then analytically continued in $a,b$.}
\begin{equation}\label{ltid}
    \int_\R\d z\;z^{-a}\,(1-z)^{-b} = \frac{2\pi\im}{(1-a-b)\,B(a,b)}\,,
\end{equation}
with $B(a,b) = \Gamma(a)\,\Gamma(b)/\Gamma(a+b)$ being the Euler beta function. The light transform is then found to be
\begin{equation}\label{scbbL}
    \bar\bL[\phi_{\Delta}^\eps](x\,|\,\zeta,\bar\zeta) = \frac{2\pi\im^{\Delta+1}\,\eps^{-\Delta}}{\Gamma(2-\Delta)}\left(\frac{\la\zeta|x|\bar\iota]}{[\bar\zeta\,\bar\iota]}\right)^{1-\Delta}\frac{1}{\la\zeta|x|\bar\zeta]-\im\,\eps\,\veps}\,,
\end{equation}
where we have reinstated the $\im\,\veps$-prescription for completeness. 
This has the expected scaling covariance in $\zeta_\al$, $\bar\zeta_{\dal}$ and is weightless in $\bar\iota_{\dal}$. Full conformal covariance is attained in the divergent limit $\bar\iota_{\dal}\to\bar\zeta_{\dal}$ as the regulator is removed.

One can also revert back to affine coordinates with a convenient choice of reference spinor. For instance, setting $\zeta_\al = (z,1)$, $\bar\zeta_{\dal} = (\bar z,1)$, and choosing $\bar\iota_{\dal} = (1,0)$, we get
\begin{equation}
    [\bar\zeta\,\bar\iota] = 1\,,\;\; \la\zeta|x|\bar\zeta] = q\cdot x\,,\;\; \partial_{\bar z}\bar\zeta_{\dal} = \bar\iota_{\dal}\,,\;\; \text{and}\;\; \la\zeta|x|\bar\iota] = \p_{\bar z}q\cdot x\,.
\end{equation} Here we have introduced the standard notation
\begin{equation}
    q_{\al\dal}(z,\bar z) = \zeta_\al\,\bar\zeta_{\dal} =  \begin{pmatrix}z\bar z&&z\\\bar z&&1\end{pmatrix}\,.
\end{equation}
So the light transform reduces to
\begin{equation}\label{scaffine}
    \bar\bL[\phi_{\Delta}^\eps](x\,|\,z,\bar z) = \frac{2\pi\im^{\Delta+1}\,\eps^{-\Delta}}{\Gamma(2-\Delta)}\,\frac{\left(\p_{\bar z}q\cdot x\right)^{1-\Delta}}{q\cdot x-\im\,\eps\,\veps}\,.
\end{equation}
This is consistent with the analogous result for $\bL[\phi_{\Delta}^\eps]$ first found in \cite{Atanasov:2021cje}. Their wavefunction can be obtained by replacing $\p_{\bar z}q\cdot x$ with $\p_zq\cdot x$ (and changing the sign conventions for the metric).


\subsection{Gluons}
\label{sec:gluonlt}

In general, wavefunctions obtained from Mellin transforming momentum eigenstates are gauge equivalent to conformal primary solutions of the linearized free field equations. For simplicity, we will light transform the latter. 

The positive and negative helicity conformal primary gluon wavefunctions are respectively given by \cite{Pasterski:2017kqt,Pasterski:2020pdk}
\begin{align}
    a^+_{\al\dal}(x\,|\,\zeta,\bar\zeta) &= (\im\,\eps)^{-\Delta}\,\Gamma(\Delta)\,\frac{x_\al{}^{\dot\beta}\,\bar\zeta_{\dot\beta}\,\bar\zeta_{\dal}}{\la\zeta|x|\bar\zeta]^{\Delta+1}}\,,\\
    a^-_{\al\dal}(x\,|\,\zeta,\bar\zeta) &= (\im\,\eps)^{-\Delta}\,\Gamma(\Delta)\,\frac{\zeta_\al\,\zeta_\beta\,x^\beta{}_{\dal}}{\la\zeta|x|\bar\zeta]^{\Delta+1}}\,,
\end{align}
where we have suppressed the $\Delta$ and $\eps$ indices. Unlike their momentum space counterparts, these wavefunctions are free of any choices of reference spinors and hence are easier to work with. They are also straightforwardly checked to possess the correct conformal covariance and satisfy the spin 1 linearized field equations. 

We can easily extend the result of the previous section to find the light transform of $a^+$ with respect to $\zeta_\al$ and $a^-$ with respect to $\bar\zeta_{\dal}$. For example, $\bL[a^+]$ reads
\begin{equation}
    \bL[a^+_{\al\dal}](x\,|\,\zeta,\bar\zeta) = (\im\,\eps)^{-\Delta}\,\Gamma(\Delta)\,x_\al{}^{\dot\beta}\,\bar\zeta_{\dot\beta}\,\bar\zeta_{\dal}\lim_{\delta\to0}\int_{\RP^1}\frac{\D\zeta'}{\la\zeta\,\zeta'\ra^{1-\Delta-\delta}}\,\frac{1}{\la\zeta'|x|\bar\zeta]^{\Delta+1}}\,\frac{1}{\la\iota\,\zeta'\ra^\delta}\,,
\end{equation}
and one can write a similar conjugate expression for $\bar\bL[a^-]$. Passing to convenient conformal frames like \eqref{zetabarsub} (see also \eqref{zetasub} below for the corresponding substitution for $\zeta'_\al$) and integrating using \eqref{ltid} yields 
\begin{align}
    \bL[a^+_{\al\dal}](x\,|\,\zeta,\bar\zeta) &= \frac{2\pi\im^{\Delta-1}\,\eps^{-\Delta}}{\Delta\,\Gamma(1-\Delta)}\left(\frac{\la\iota|x|\bar\zeta]}{\la\zeta\,\iota\ra}\right)^{-\Delta}\frac{x_\al{}^{\dot\beta}\,\bar\zeta_{\dot\beta}\,\bar\zeta_{\dal}}{\la\zeta|x|\bar\zeta]}\,,\label{bLa+}\\
    \bar\bL[a^-_{\al\dal}](x\,|\,\zeta,\bar\zeta) &= \frac{2\pi\im^{\Delta-1}\,\eps^{-\Delta}}{\Delta\,\Gamma(1-\Delta)}\left(\frac{\la\zeta|x|\bar\iota]}{[\bar\zeta\,\bar\iota]}\right)^{-\Delta}\frac{\zeta_\al\,\zeta_\beta\,x^\beta{}_{\dal}}{\la\zeta|x|\bar\zeta]}\,.\label{bbLa-}
\end{align}
Again, $\iota_\al$ and $\bar\iota_{\dal}$ are reference spinors needed to regulate the light ray integrals. As in \eqref{scaffine}, one can easily go to the affine versions of these expressions with the standard choice $\iota_\al=(1,0)$, $\bar\iota_{\dal} = (1,0)$. The scaling weights of these states are easily verified to be the anticipated values $(\frac{1-\Delta}{2},\frac{\Delta-1}{2})$ and $(\frac{\Delta-1}{2},\frac{1-\Delta}{2})$ respectively.

The other two cases are only marginally more involved. For example, consider
\begin{equation}
    \bL[a^-_{\al\dal}](x\,|\,\zeta,\bar\zeta) = (\im\,\eps)^{-\Delta}\,\Gamma(\Delta)\lim_{\delta\to0}\int_{\RP^1}\frac{\D\zeta'}{\la\zeta\,\zeta'\ra^{3-\Delta-\delta}}\,\frac{\zeta'_\al\,\zeta'_\beta\,x^\beta{}_{\dal}}{\la\zeta'|x|\bar\zeta]^{\Delta+1}}\,\frac{1}{\la\iota\,\zeta'\ra^\delta}
\end{equation}
with $\iota_\al$ a reference spinor. As before, the integrand is invariant under rescaling $\zeta'_\al$. We go to the conformal frame
\begin{equation}\label{zetasub}
    \zeta'_\al = \zeta_\al - z'\,\frac{\la\zeta|x|\bar\zeta]}{\la\iota|x|\bar\zeta]}\,\iota_\al
\end{equation}
in which the points $\zeta_\al$, $x_\al{}^{\dal}\bar\zeta_{\dal}$ and $\iota_\al$ are mapped to $z'=0,1,\infty$ respectively. This leaves us with a bunch of integrals over $z'$:
\begin{multline}
    (\im\,\eps)^{-\Delta}\,\Gamma(\Delta)\left(\frac{\la\iota|x|\bar\zeta]}{\la\iota\,\zeta\ra}\right)^{2-\Delta}\frac{x^\beta{}_{\dal}}{\la\zeta|x|\bar\zeta]^3}\,\lim_{\delta\to0}\int_\R\d z'\;z'^{\Delta-3-\delta}\,(1-z')^{-1-\Delta}\\
    \times\left(\zeta_\al\,\zeta_\beta-2\,z'\,\frac{\la\zeta|x|\bar\zeta]}{\la\iota|x|\bar\zeta]}\,\zeta_{(\al}\,\iota_{\beta)} + z'^2\,\frac{\la\zeta|x|\bar\zeta]^2}{\la\iota|x|\bar\zeta]^2}\,\iota_\al\,\iota_\beta\right)\,.
\end{multline}
These are again the same kind of integrals that we encountered earlier in \eqref{ltid}. They finally yield
\begin{multline}
    \bL[a^-_{\al\dal}](x\,|\,\zeta,\bar\zeta) = \frac{2\pi\im^{\Delta-1}\,\eps^{-\Delta}}{\Delta\,\Gamma(3-\Delta)}\left(\frac{\la\iota|x|\bar\zeta]}{\la\iota\,\zeta\ra}\right)^{2-\Delta}\frac{x^\beta{}_{\dal}}{\la\zeta|x|\bar\zeta]^3}\\
    \times\left(2\,\zeta_\al\,\zeta_\beta+2\,(\Delta-2)\,\frac{\la\zeta|x|\bar\zeta]}{\la\iota|x|\bar\zeta]}\,\iota_{(\al}\,\zeta_{\beta)} + (\Delta-1)\,(\Delta-2)\,\frac{\la\zeta|x|\bar\zeta]^2}{\la\iota|x|\bar\zeta]^2}\,\iota_\al\,\iota_\beta\right)\,.
\end{multline}
This has conformal weights $(\frac{3-\Delta}{2},\frac{\Delta+1}{2})$, and has weight $0$ under scalings of $\iota_\al$ as expected when the regulator $\delta\to0$. Similarly, using substitutions of the kind \eqref{zetabarsub}, we find
\begin{multline}
    \bar\bL[a^+_{\al\dal}](x\,|\,\zeta,\bar\zeta) = \frac{2\pi\im^{\Delta-1}\,\eps^{-\Delta}}{\Delta\,\Gamma(3-\Delta)}\left(\frac{\la\zeta|x|\bar\iota]}{[\bar\zeta\,\bar\iota]}\right)^{2-\Delta}\frac{x_\al{}^{\dot\beta}}{\la\zeta|x|\bar\zeta]^3}\\
    \times\left(2\,\bar\zeta_{\dal}\,\bar\zeta_{\dot\beta}+2\,(\Delta-2)\,\frac{\la\zeta|x|\bar\zeta]}{\la\zeta|x|\bar\iota]}\,\bar\iota_{(\dal}\,\bar\zeta_{\dot\beta)} + (\Delta-1)\,(\Delta-2)\,\frac{\la\zeta|x|\bar\zeta]^2}{\la\zeta|x|\bar\iota]^2}\,\bar\iota_{\dal}\,\bar\iota_{\dot\beta}\right)\,,
\end{multline}
which has weights $(\frac{\Delta+1}{2},\frac{3-\Delta}{2})$. 


\subsection{Gravitons}

The conformal primary graviton wavefunctions \cite{Pasterski:2017kqt,Pasterski:2020pdk} take a similar form,
\begin{align}
    h^+_{\al_1\dal_1\al_2\dal_2}(x\,|\,\zeta,\bar\zeta) &= (\im\,\eps)^{-\Delta}\,\Gamma(\Delta)\,\frac{x_{\al_1}{}^{\dal_3}\,x_{\al_2}{}^{\dal_4}\,\bar\zeta_{\dal_1}\,\bar\zeta_{\dal_2}\,\bar\zeta_{\dal_3}\,\bar\zeta_{\dal_4}}{\la\zeta|x|\bar\zeta]^{\Delta+2}}\,,\\
    h^-_{\al_1\dal_1\al_2\dal_2}(x\,|\,\zeta,\bar\zeta) &= (\im\,\eps)^{-\Delta}\,\Gamma(\Delta)\,\frac{\zeta_{\al_1}\,\zeta_{\al_2}\,\zeta_{\al_3}\,\zeta_{\al_4}\,x^{\al_3}{}_{\dal_1}\,x^{\al_4}{}_{\dal_2}}{\la\zeta|x|\bar\zeta]^{\Delta+2}}\,.
\end{align}
Computing the easier pair of light transforms yields
\begin{multline}\label{bLh+}
    \bL[h^+_{\al_1\dal_1\al_2\dal_2}](x\,|\,\zeta,\bar\zeta) = \frac{2\pi\im^{\Delta-3}\,\eps^{-\Delta}}{\Delta\,(\Delta+1)\,\Gamma(-\Delta)}\\\times\left(\frac{\la\iota|x|\bar\zeta]}{\la\zeta\,\iota\ra}\right)^{-\Delta-1}\frac{x_{\al_1}{}^{\dal_3}\,x_{\al_2}{}^{\dal_4}\,\bar\zeta_{\dal_1}\,\bar\zeta_{\dal_2}\,\bar\zeta_{\dal_3}\,\bar\zeta_{\dal_4}}{\la\zeta|x|\bar\zeta]}\,,
\end{multline}
and
\begin{multline}\label{bbLh-}
    \bar\bL[h^-_{\al_1\dal_1\al_2\dal_2}](x\,|\,\zeta,\bar\zeta) = \frac{2\pi\im^{\Delta-3}\,\eps^{-\Delta}}{\Delta\,(\Delta+1)\,\Gamma(-\Delta)}\\\times\left(\frac{\la\zeta|x|\bar\iota]}{[\bar\zeta\,\bar\iota]}\right)^{-\Delta-1}\frac{\zeta_{\al_1}\,\zeta_{\al_2}\,\zeta_{\al_3}\,\zeta_{\al_4}\,x^{\al_3}{}_{\dal_1}\,x^{\al_4}{}_{\dal_2}}{\la\zeta|x|\bar\zeta]}\,.
\end{multline}
The former has conformal weights $(-\frac{\Delta}{2},\frac{\Delta-2}{2})$, while the latter has weights $(\frac{\Delta-2}{2},-\frac{\Delta}{2})$.

On the other hand, the wavefunction for $\bL[h^-]$ is given by
\begin{multline}
    \bL[h^-_{\al_1\dal_1\al_2\dal_2}](x\,|\,\zeta,\bar\zeta) =(\im\,\eps)^{-\Delta}\,\Gamma(\Delta)\lim_{\delta\to0}\int_{\RP^1}\frac{\D\zeta'}{\la\zeta\,\zeta'\ra^{4-\Delta-\delta}}\,\frac{1}{\la\iota\,\zeta'\ra^\delta}\\
    \times \frac{x^{\al_3}{}_{\dal_1}\,x^{\al_4}{}_{\dal_2}}{\la\zeta'|x|\bar\zeta]^{\Delta+2}}\prod_{i=1}^4\zeta'_{\al_i}\,.
\end{multline}
This can be evaluated using the same substitution \eqref{zetasub} as before. Using the binomial expansion
\begin{equation}
    \prod_{i=1}^4\zeta'_{\al_i} = \sum_{r=0}^4{4\choose r}\left(- z'\,\frac{\la\zeta|x|\bar\zeta]}{\la\iota|x|\bar\zeta]}\right)^r \iota_{(\al_1}\cdots\iota_{\al_r}\zeta_{\al_{r+1}}\cdots\zeta_{\al_4)}\,,
\end{equation}
the integral becomes
\begin{multline}
    (\im\,\eps)^{-\Delta}\,\Gamma(\Delta)\left(\frac{\la\iota|x|\bar\zeta]}{\la\iota\,\zeta\ra}\right)^{3-\Delta}\frac{x^{\al_3}{}_{\dal_1}\,x^{\al_4}{}_{\dal_2}}{\la\zeta|x|\bar\zeta]^5}\sum_{r=0}^4{4\choose r}\left(-\frac{\la\zeta|x|\bar\zeta]}{\la\iota|x|\bar\zeta]}\right)^r \iota_{(\al_1}\cdots\iota_{\al_r}\zeta_{\al_{r+1}}\cdots\zeta_{\al_4)}\\
    \times\lim_{\delta\to0}\int_\R\d z'\;z'^{\Delta-4+r-\delta}\,(1-z')^{-2-\Delta}\,.
\end{multline}
Applying \eqref{ltid} yields the wavefunction
\begin{multline}
     \bL[h^-_{\al_1\dal_1\al_2\dal_2}](x\,|\,\zeta,\bar\zeta) = \frac{2\pi\im^{\Delta-3}\,\eps^{-\Delta}}{\Delta\,(\Delta+1)\,\Gamma(4-\Delta)}\left(\frac{\la\iota|x|\bar\zeta]}{\la\zeta\,\iota\ra}\right)^{3-\Delta}\frac{x^{\al_3}{}_{\dal_1}\,x^{\al_4}{}_{\dal_2}}{\la\zeta|x|\bar\zeta]^5}\\
     \times\sum_{r=0}^4\frac{4!\,(\Delta-3)_r}{r!}\,\frac{\la\zeta|x|\bar\zeta]^r}{\la\iota|x|\bar\zeta]^r}\,\iota_{(\al_1}\cdots\iota_{\al_r}\zeta_{\al_{r+1}}\cdots\zeta_{\al_4)}\,.
\end{multline}
In the second line, we have used the notation $(\Delta-3)_r = \Gamma(\Delta-3+r)/\Gamma(\Delta-3)$ for the Pochhammer symbol. The conformal weights of this state are seen to be $(\frac{4-\Delta}{2},\frac{\Delta+2}{2})$. Similarly, we find the conjugate expression,
\begin{multline}
     \bar\bL[h^+_{\al_1\dal_1\al_2\dal_2}](x\,|\,\zeta,\bar\zeta) = \frac{2\pi\im^{\Delta-3}\,\eps^{-\Delta}}{\Delta\,(\Delta+1)\,\Gamma(4-\Delta)}\left(\frac{\la\zeta|x|\bar\iota]}{[\bar\zeta\,\bar\iota]}\right)^{3-\Delta}\frac{x_{\al_1}{}^{\dal_3}\,x_{\al_2}{}^{\dal_4}}{\la\zeta|x|\bar\zeta]^5}\\
     \times\sum_{r=0}^4\frac{4!\,(\Delta-3)_r}{r!}\,\frac{\la\zeta|x|\bar\zeta]^r}{\la\zeta|x|\bar\iota]^r}\,\bar\iota_{(\dal_1}\cdots\bar\iota_{\dal_r}\bar\zeta_{\dal_{r+1}}\cdots\bar\zeta_{\dal_4)}\,,
\end{multline}
with weights $(\frac{\Delta+2}{2},\frac{4-\Delta}{2})$. Of course, the same considerations would also apply to higher integer spins.


\section{Transform to twistor space}
\label{sec:twistor}

With the definition of light transforms in hand, we can try defining a half-Fourier transform of boost eigenstates and comparing the two notions. We will define primary twistor eigenstates and show how precisely they are equivalent to a linear combination of positive and negative frequency light transformed boost eigenstates. This section mainly serves as a motivation for the next, and the reader may easily find it summarized in the commutative diagram \eqref{comd} and the paragraph enclosing it.

\subsection{Half-Mellin transform}

Start again with a massless, spin $\ell$ momentum eigenstate $|\lambda,\bar\lambda,\ell\ra$. Its Mellin transform was defined in \eqref{cpstate} by fixing its little group scaling. However, working more invariantly, we can consider keeping this redundancy in play.\footnote{For example, this freedom was used in \cite{Brandhuber:2021nez} in Minkowski space to define a ``chiral Mellin transform''. Our ``half-Mellin transform'' is a somewhat analogous construction in split signature.} The state is then labeled by four parameters $t,\bar t,z,\bar z$ as in \eqref{shvar}. We define a half-Mellin transform as a Mellin transform in the absolute value of either $t$ or $\bar t$.

The transform along $t$ is defined by first breaking up $\lambda_\al = \eps\,t\,\zeta_\al$ with $\eps\in\{\pm1\}$ and $t\in\R_+$. We then Mellin transform in $t$ weighted by a factor of $t^{2h}$ instead of $t^\Delta$:
\begin{equation}\label{tmellin}
    |\zeta,\bar\lambda,h,\ell,\eps\ra \vcentcolon = \int_{\R_+}\frac{\d t}{t}\,t^{2h}\,|\eps\,t\,\zeta,\bar\lambda,\ell\ra\,.
\end{equation}
Though we have labeled the transformed state by $h = (\Delta+\ell)/2$ and $\ell$ but not $\bar h = (\Delta-\ell)/2$, it is actually covariant under conformal rescalings of $\zeta_\al$, $\bar\lambda_{\dal}$ with the same weights $h,\bar h$ as the good old boost eigenstate $|\zeta,\bar\zeta,h,\bar h,\eps\ra$,
\begin{equation}
    |s\,\zeta,\bar s\,\bar\lambda,h,\ell,\eps\ra = s^{-2h}\,\bar s^{-2\bar h}\,|\zeta,\bar\lambda,h,\eps,\ell\ra\,,\qquad s,\bar s\in\R_+\,.
\end{equation}
This is true because the two states are actually related. If we fix the little group scaling to set $\bar\lambda_{\dal} = (\bar z,1)\equiv\bar\zeta_{\dal}$, and use little group covariance \eqref{lg} in the form
\begin{equation}
    |\eps\,t\,\zeta,\bar\zeta,\ell\ra = (\eps\,\sqrt{t})^{-2\ell}\,|\sqrt{t}\,\zeta,\eps\,\sqrt{t}\,\bar\zeta,\ell\ra\,,
\end{equation}
we can convert \eqref{tmellin} into
\begin{equation}
     |\zeta,\bar\lambda=\bar\zeta,h,\ell,\eps\ra = \int_{\R_+}\frac{\d t}{t}\,t^{\Delta}\,\eps^{-2\ell}\,|\sqrt{t}\,\zeta,\eps\,\sqrt{t}\,\bar\zeta,\ell\ra = \eps^{-2\ell}\,|\zeta,\bar\zeta,h,\bar h,\eps\ra\,.
\end{equation}
Thus, the half-Mellin transformed state is equivalent to the standard boost eigenstate for $\bar\lambda_{\dal} = (\bar z,1)$. But it provides a convenient generalization of the boost eigenstate when we do not wish to fix the scale of $\bar\lambda_{\dal}$ in this way, especially when we would like to Fourier transform independently in its two components in what follows.

Similarly, we can define the half-Mellin transform in the magnitude of $\bar t$ while keeping $\lambda_\al$ arbitrary. Simply break up $\bar\lambda_{\dal} = \eps\,\bar t\,\bar\zeta_{\dal}$ with $\eps$ its sign, and Mellin transform in $\bar t$:
\begin{equation}
    |\lambda,\bar\zeta,\bar h,\ell,\eps\ra \vcentcolon = \int_{\R_+}\frac{\d\bar t}{\bar t}\,\bar t^{2\bar h}\,|\lambda,\eps\,\bar t\,\bar\zeta,\ell\ra\,.
\end{equation}
When $\lambda_\al = (z,1)\equiv\zeta_\al$, we again land on the boost eigenstate,
\begin{equation}
    |\lambda=\zeta,\bar\zeta,\bar h,\ell,\eps\ra = |\zeta,\bar\zeta,h,\bar h,\eps\ra\,.
\end{equation}
We can immediately put these definitions to use by transforming these ``generalized'' boost eigenstates to twistor space.


\subsection{Half-Fourier transform}

Witten's ``half-Fourier'' transform takes momentum eigenstates $|\lambda,\bar\lambda,\ell\ra$ as input and defines a state labeled by a twistor $Z^A = (\lambda_\al,\mu^{\dal})\in\RP^3$,
\begin{equation}\label{Zstate}
    |Z,\ell\ra_\T \equiv |\lambda,\mu,\ell\ra_\T \vcentcolon = \int_{\R^2}\d^2\bar\lambda\;\e^{\im[\mu\,\bar\lambda]}\,|\lambda,\bar\lambda,\ell\ra\,.
\end{equation}
We have labeled the state further by a subscript $\T$ to denote that it has been obtained via such a Fourier transform. Under a scaling $Z\mapsto r\,Z$, it transforms covariantly:
\begin{equation}\label{twistorscale}
    |r\,Z,\ell\ra_\T = r^{-2\ell-2}\,|Z,\ell\ra_\T\,,\qquad r\in\R^*\,.
\end{equation}
Analogously, one can also Fourier transform in $\lambda_\al$ to get a state labeled by a dual twistor $W_A = (\bar\mu^\al,\bar\lambda_{\dal})\in\RP^3$,
\begin{equation}
    |W,\ell\ra_{\bar\T}\equiv |\bar\mu,\bar\lambda,\ell\ra_{\bar\T} \vcentcolon = \int_{\R^2}\d^2\lambda\;\e^{\im\la\bar\mu\,\lambda\ra}\,|\lambda,\bar\lambda,\ell\ra\,.
\end{equation}
Its transformation under scalings is similar:
\begin{equation}
    |r\,W,\ell\ra_{\bar\T} = r^{2\ell-2}\,|W,\ell\ra_{\bar\T}\,,\qquad r\in\R^*\,.
\end{equation}
Scattering of such states was studied by Witten in \cite{Witten:2003nn} and explored in greater detail in \cite{ArkaniHamed:2009si, Mason:2009sa}.

We would now like to define the analogues of conformal primary states for these twistor/dual twistor eigenstates. These can be set up by half-Fourier transforming the half-Mellin transformed states of the previous subsection. For instance, writing $\sZ^A = (\zeta_\al,\mu^{\dal})$, define the \emph{conformal primary twistor eigenstate}
\begin{equation}\label{tmstate}
    |\sZ,h,1-\bar h,\eps\ra_\T \equiv |\zeta,\mu,h,1-\bar h,\eps\ra_\T \vcentcolon = \int_{\R^2}\d^2\bar\lambda\;\e^{\im[\mu\,\bar\lambda]}\,|\zeta,\bar\lambda,h,\ell,\eps\ra\,.
\end{equation}
We can also insert the expression \eqref{tmellin} of the half-Mellin transform here, exchange the order of integrals, and write this as a half-Mellin transform of the twistor eigenstate
\begin{equation}\label{tmstate1}
    |\zeta,\mu,h,1-\bar h,\eps\ra_\T = \int_{\R^+}\frac{\d t}{t}\,t^{2h}\,|\eps\,t\,\zeta,\mu,\ell\ra_\T\,.
\end{equation}
Thus, \emph{the primary twistor eigenstate is a half-Mellin transform of the twistor eigenstate.} This is useful in actual computations.

Applying \eqref{twistorscale}, one sees that \eqref{tmstate1} satisfies the conformal transformation law that its quantum numbers indicate:
\begin{equation}\label{tmscale}
    |s\,\zeta,\bar s\,\mu,h,1-\bar h,\eps\ra_\T = s^{-2h}\,\bar s^{-2(1-\bar h)}\,|\zeta,\mu,h,1-\bar h,\eps\ra_\T\,,\qquad s,\bar s\in\R_+\,.
\end{equation}
Since the conformal rescalings of $\mu^{\dal}$ and $\zeta_\al$ have completely decoupled, we can view them as living in separate copies of $\RP^1$. So, if we allow ourselves to think of $(\zeta_\al,\mu_{\dal})$ not as a twistor but instead as coordinates on another copy of the celestial torus, we see that this half-Mellin transform has generated for us a conformal primary state with weights $(h,1-\bar h)$. These are the same weights as those of the light transformed state $|\zeta,\bar\zeta,h,1-\bar h,\eps\ra_{\bar\bL}$ given in \eqref{Lbstate}, raising the possibility of expressing one in terms of the other! We will see in the next subsection that this can indeed be done.

Similarly, setting $\sW_A = (\bar\mu^\al,\bar\zeta_{\dal})$ and defining
\begin{equation}\label{dtmstate}
    \begin{split}
        |\sW,1-h,\bar h,\eps\ra_{\bar\T} \equiv |\bar\mu,\bar\zeta,1-h,\bar h,\eps\ra_{\bar\T} &\vcentcolon = \int_{\R^2}\d^2\lambda\;\e^{\im\la\bar\mu\,\lambda\ra}\,|\lambda,\bar\zeta,\bar h,\ell,\eps\ra\\
        &= \int_{\R_+}\frac{\d\bar t}{\bar t}\,\bar t^{2\bar h}\,|\bar\mu,\eps\,\bar t\,\bar\zeta,\ell\ra_{\bar\T}\,,
    \end{split}
\end{equation}
we observe that
\begin{equation}\label{dtmscale}
    |s\,\bar\mu,\bar s\,\bar\zeta,h,\bar h,\eps\ra_{\bar\T} = s^{-2(1-h)}\,\bar s^{-2\bar h}\,|\bar\mu,\bar\zeta,1-h,\bar h,\eps\ra_{\bar\T}\,,\qquad s,\bar s\in\R_+\,.
\end{equation}
As anticipated by now, this has the same conformal weights in $(\bar\mu^\al,\bar\zeta_{\dal})$ as the other light transformed state $|\zeta,\bar\zeta,1-h,\bar h,\eps\ra_\bL$ had in $(\zeta_\al,\bar\zeta_{\dal})$.

We remark in passing that $\sZ^A=(\mu^{\dal},\zeta_\al)$ can be viewed as a twistor for the AdS$_3/\Z$ slices that foliate $\R^{2,2}$ and are mapped to each other under dilatations \cite{Atanasov:2021oyu}. This is an example of symmetry reduction wherein one quotients the twistor space $\RP^3$ by the dilatation vector field $\mu^{\dal}\,\p_{\mu^{\dal}}$ (after removing the fixed points of the dilatation). This yields $\RP^1\times\RP^1$ as the twistor space for AdS$_3/\Z$ \cite{Jones:1985pla}. Slices of $\RP^1\times\RP^1$ then ``foliate'' $\RP^3$. The $\RP^1$ containing $\mu^{\dal}$ can be thought of as being ``Fourier-dual'' to the $\bar\zeta_{\dal}\in\RP^1$ factor of the celestial torus. Similarly, the space of $\sW_A$'s provides a dual twistor space. 


\subsection{Half-Fourier transform $\cong$ light transform}
\label{sec:hfislight}

To see how the twistorial and light transforms are related, we will expand the state $|\sZ,h,\bar h,\eps\ra_\T$ given by \eqref{tmstate} in terms of the boost eigenstates of \eqref{cpstate}. 

We begin with the convolution of Mellin and Fourier integrals that follows from combining \eqref{Zstate} and \eqref{tmstate},
\begin{equation}\label{momtoZ}
    |\sZ,h,1-\bar h,\eps\ra_\T = \int_{\R^2}\d^2\bar\lambda\;\e^{\im[\mu\,\bar\lambda]}\int_{\R_+}\frac{\d t}{t}\,t^{2h}\,|\eps\,t\,\zeta,\bar\lambda,\ell\ra\,.
\end{equation}
As in \eqref{shvar}, we can decompose the integration variable
\begin{equation}
    \bar\lambda_{\dal} = \bar t\,\bar\zeta_{\dal} = \bar\eps\,|\bar t|\,\bar\zeta_{\dal}\,,\qquad\bar\eps = \sgn\,\bar t\,,
\end{equation}
with $\bar\zeta_{\dal} = (\bar z,1)$. The integration measure splits as $\d^2\bar\lambda = \bar t\,\d\bar t\,\D\bar\zeta = \bar t\,\d\bar t\,\d\bar z$. Then we can use little group scaling \eqref{lg} to write
\begin{equation}
    |\eps\,t\,\zeta,\bar\lambda,\ell\ra = |\eps\,t\,\zeta,\bar\eps\,|\bar t|\,\bar\zeta,\ell\ra = \left(\eps\,\sqrt{\frac{t}{|\bar t|}}\,\right)^{-2\ell}\bigl|\sqrt{t\,|\bar t|}\,\zeta,\eps\,\bar\eps\sqrt{t\,|\bar t|}\,\bar\zeta,\ell\bigr\ra\,.
\end{equation}
Consequently, \eqref{momtoZ} becomes
\begin{equation}\label{momtoZ1}
    |\sZ,h,1-\bar h,\eps\ra_\T = \eps^{-2\ell}\int_{\RP^1}\D\bar\zeta\int_{\R^*}\frac{\d\bar t}{\bar t}\,|\bar t|^{2+\ell}\,\e^{\im\bar t[\mu\,\bar\zeta]}\int_{\R_+}\frac{\d t}{t}\,t^{\Delta}\,\bigl|\sqrt{t\,|\bar t|}\,\zeta,\eps\,\bar\eps\sqrt{t\,|\bar t|}\,\bar\zeta,\ell\bigr\ra\,,
\end{equation}
having noted that $t^{2h}\,t^{-\ell}=t^\Delta$. Substituting $t = \omega/|\bar t|$, $\omega\in\R_+$, the $\omega$ integral reduces to the definition \eqref{cpstate} of a boost eigenstate. This leaves us with
\begin{equation}\label{momtoZ2}
    |\sZ,h,1-\bar h,\eps\ra_\T = \eps^{-2\ell}\int_{\RP^1}\D\bar\zeta\int_{\R^*}\frac{\d\bar t}{\bar t}\,|\bar t|^{2-2\bar h}\,\e^{\im\bar t[\mu\,\bar\zeta]}\,|\zeta,\bar\zeta,h,\bar h,\eps\,\bar\eps\ra\,,
\end{equation}
where we have used $|\bar t|^{2+\ell}|\bar t|^{-\Delta} = |\bar t|^{2-2\bar h}$.

To perform the $\bar t$ integral, we can split it over the ranges $\bar t>0$ and $\bar t<0$. Applying
\begin{equation}
    \int_{\R_+}\frac{\d\bar t}{\bar t}\,\bar t^{2-2\bar h}\,\e^{\im\bar t[\mu\,\bar\zeta]-\veps\,\bar t} = \frac{\im^{2-2\bar h}\,\Gamma(2-2\bar h)}{([\mu\,\bar\zeta]+\im\,\veps)^{2-2\bar h}}
\end{equation}
and sending the regulator $\veps\to 0^+$, we find an expansion with two terms:
\begin{equation}\label{momtoZ3}
    |\sZ,h,1-\bar h,\eps\ra_\T = \eps^{-2\ell}\,\im^{2-2\bar h}\,\Gamma(2-2\bar h)\sum_{\bar\eps=\pm1}\bar\eps^{2\bar h-1}\int_{\RP^1}\frac{\D\bar\zeta}{[\mu\,\bar\zeta]^{2-2\bar h}}\,|\zeta,\bar\zeta,h,\bar h,\eps\,\bar\eps\ra\,.
\end{equation}
Comparing this with \eqref{Lbstate}, we recognize this to be a combination of light transformed states evaluated at the point $(\zeta_\al,\mu_{\dal})\in\RP^1\times\RP^1$ on the physical celestial torus,
\begin{equation}\label{momtoZ4}
    |\zeta,\mu,h,1-\bar h,\eps\ra_\T = \im^{2-2\bar h}\,\Gamma(2-2\bar h)\sum_{\bar\eps=\pm\eps}\eps^{1-2h}\,\bar\eps^{2\bar h-1}\,|\zeta,\mu,h,1-\bar h,\bar\eps\ra_{\bar\bL}\,.
\end{equation}
In the last step, we redefined $\bar\eps$ by shifting $\bar\eps\to\bar\eps/\eps$. For clarity, we have labeled both sides with $\zeta_\al,\mu_{\dal}$ instead of with twistors.\footnote{This is where the subscripts $\T$ and $\bL$ really come in handy.}

Thus, we conclude that the following diagram of integral transforms commutes: 
\begin{equation}\label{comd}
    \begin{tikzcd}[column sep = 60pt, row sep = 50pt]
        \substack{\text{Momentum}\\\text{eigenstates}}\arrow[r, "\substack{\frac{1}{2}\text{-Fourier}\\\text{transform}}"]\arrow[d,"\substack{\text{Mellin}\\\text{transform}}"']& \substack{\text{Twistor}\\\text{ eigenstates}}\arrow[d,"\substack{\frac{1}{2}\text{-Mellin}\\\text{transform}}"]\\
        \substack{\text{Boost}\\\text{eigenstates}}\arrow[r, "\substack{\text{Light}\\\text{transforms}}"]& \substack{\text{Primary twistor}\\\text{eigenstates}}
    \end{tikzcd}
\end{equation}
with the understanding that the bottom arrow labeled ``light transforms'' includes taking the two-term linear combination instructed by \eqref{momtoZ4}. Hence, Mellin transforms on both sides provide a natural map between the two notions of twistor eigenstates and light transformed boost eigenstates.

\section{Light transformed amplitudes}
\label{sec:ltamp}

\subsection{Choice of ambidextrous basis}

The main idea driving the discovery of the link/Grassmannian representation of gluon amplitudes in \cite{ArkaniHamed:2009si} was to transform them to twistor/dual twistor space depending on helicity. The authors there chose to transform negative helicity states to twistor eigenstates by Fourier transforming in their $\bar\lambda_{\dal}$ variables, and similarly transformed positive helicity states to dual twistor space by Fourier transforming the $\lambda_\al$'s. This was referred to as an ambidextrous basis of states.

We have seen in the previous section that in the celestial context, the light transform can be considered as the natural analogue of half-Fourier transforms. 
In fact, it can even be thought of as being more fundamental than the latter as it ``refines'' twistor eigenstates into a two-term linear combination. Though we won't go into the computational details of celestial amplitudes of primary twistor eigenstates, this refinement is morally the reason why light transformed celestial amplitudes will not display the distributional characteristics of their conformal primary cousins. Taking inspiration from the success of half-Fourier transforms in uncovering both stringy and Grassmannian geometries of amplitudes, we can try working with the corresponding ambidextrous basis of light transformed states.\footnote{It is perhaps interesting to also try working directly with primary twistor eigenstates. Unfortunately, preliminary results indicate that their celestial amplitudes are again distributional at low multiplicity.}

As we saw in section \ref{sec:gluonlt}, it is much more natural to light transform positive and negative helicity particles respectively in their $\zeta_\al$ and $\bar\zeta_{\dal}$ (i.e., $z$ and $\bar z$) celestial positions than the other way round. This is in line with the choice of the aforementioned ambidextrous basis of twistor eigenstates. So, we choose the spin 1 states $\bL[a^+]$, $\bar\bL[a^-]$ given by \eqref{bLa+}, \eqref{bbLa-} as our ambidextrous basis of light transformed gluons, and the corresponding spin 2 states $\bL[h^+]$, $\bar\bL[h^-]$ from \eqref{bLh+}, \eqref{bbLh-} for gravitons. Given a celestial amplitude $\cA_n$ with positive and negative helicity gluons/gravitons indexed by the symbols $\sa$ and $\bsa$ respectively, we will look at the following light transformed celestial amplitude:
\begin{multline}\label{Ldef0}
    \cL_n(z_i,\bar z_i,\Delta_i,\ell_i,\eps_i) \defeq \int_{\R^n}\prod_{\bsa}\frac{\d\bar z'_{\bsa}}{(\bar z'_{\bsa}-\bar z_{\bsa})^{2-2\bar h_{\bsa}}}\prod_\sa\frac{\d z'_\sa}{(z'_\sa-z_\sa)^{2-2h_\sa}}\\
    \times\cA_n(z_{\bsa},\bar z'_{\bsa},\Delta_{\bsa},-,\eps_{\bsa}\,;\,z'_\sa,\bar z_\sa,\Delta_\sa,+,\eps_\sa)\,.
\end{multline}
We can also express this in homogeneous coordinates,
\begin{multline}\label{Ldef}
    \cL_n(\zeta_i,\bar\zeta_i,\Delta_i,\ell_i,\eps_i) \defeq \int_{(\RP^1)^n}\prod_{\bsa}\frac{\D\bar\zeta'_{\bsa}}{[\bsa\,\bsa']^{2-2\bar h_{\bsa}}}\prod_\sa\frac{\D\zeta'_\sa}{\la\sa\,\sa'\ra^{2-2h_\sa}}\\
    \times\cA_n(\zeta_{\bsa},\bar\zeta'_{\bsa},\Delta_{\bsa},-,\eps_{\bsa}\,;\,\zeta'_\sa,\bar\zeta_\sa,\Delta_\sa,+,\eps_\sa)\,,
\end{multline}
having used the abbreviations
\begin{equation}
    [\bar\zeta_{\bsa}\,\bar\zeta'_{\bsa}] \equiv [\bsa\,\bsa']\,,\qquad \la\zeta_\sa\,\zeta'_\sa\ra \equiv \la\sa\,\sa'\ra\,,\;\text{etc.}
\end{equation}
There is no fear of confusing these with standard spinor-helicity contractions like $[i\,j] = [\bar\lambda_i\,\bar\lambda_j]$, $\la i\,j\ra = \la\lambda_i\,\lambda_j\ra$, etc., as we will never use the latter.

The construction of $\cL_n$ treats all gluons of a given helicity on equal footing. This is to be contrasted with the procedure of singling out a specific gluon to be shadow transformed as in \cite{Fan:2021isc,Crawley:2021ivb}. We can further simplify the definition of $\cL_n$ by recalling that with our helicity choices: $2-2\bar h_{\bsa} = 1-\Delta_{\bsa}$ and $2-2h_\sa = 1-\Delta_\sa$ for gluons, and $2-2\bar h_{\bsa} = -\Delta_{\bsa}$ and $2-2h_\sa = -\Delta_\sa$ for gravitons. This also puts the two kinds of light transforms on equal footing.


\subsection{Examples of gluon amplitudes}
\label{sec:gluonltamp}

To build intuition, let us study light transformed gluon amplitudes in the simple cases of 2 and 3 points. For ease of comparison, the calculations in this section will be done using the usual affine celestial coordinates.  As a matter of notation, $n$-point celestial and light transformed amplitudes will be concisely written as
\begin{equation}
    \begin{split}
        &\cA_n(1^{\eps_1}_{\Delta_1,\ell_1},2^{\eps_2}_{\Delta_2,\ell_2},\dots,n^{\eps_n}_{\Delta_n,\ell_n})\,,\\
        &\cL_n(1^{\eps_1}_{\Delta_1,\ell_1},2^{\eps_2}_{\Delta_2,\ell_2},\dots,n^{\eps_n}_{\Delta_n,\ell_n})\,.
    \end{split}
\end{equation}
Any such amplitude will contain a factor of $\delta(\beta)$, with $\beta$ standing for the quantity
\begin{equation}
    \beta = \im\sum_{i=1}^n(\Delta_i-1)\,.
\end{equation}
This is the conserved charge associated to overall boost invariance. To make sense of $\delta(\beta)$, one generally assumes that the $\Delta_i$ lie on the principal series $1+\im\,\R.$

\paragraph{2 points.} 2-point functions in CCFT are defined to be the Mellin transforms of inner products of bulk momentum eigenstates. The 2-gluon celestial amplitude/inner product is given by \cite{Pasterski:2017ylz}
\begin{equation}\label{cA2gluon}
    \cA_2(1_{\Delta_1,\ell}^{\eps},2_{\Delta_2,-\ell}^{-\eps}) = 2\pi\,\cC^\eps_{\Delta_1,\ell}\,\delta(\im(\Delta_1+\Delta_2-2))\,\delta(z_{12})\,\delta(\bar z_{12})\,,
\end{equation}
where $z_{ij}\equiv z_i-z_j$, etc. We have kept a factor $\cC_{\Delta,\ell}^\eps$ denoting a normalization that one may fix depending on application. Since we need $\Delta_2=2-\Delta_1$, $\ell_2=-\ell_1$ and $\eps_2=-\eps_1$ to get a non-vanishing 2-point function, the normalization only needs to depend on one of the particles' weights. We have chosen this to be the first particle.

Without loss of generality, we can take the first gluon to be outgoing and positive helicity, i.e., $\eps=+1$, $\ell=+1$. Following \eqref{Ldef0}, we light transform in $z_1$ and $\bar z_2$:
\begin{equation}\label{L2gluon}
    \begin{split}
        \cL_2(1_{\Delta_1,+1}^{+},2_{\Delta_2,-1}^{-}) &= 2\pi\,\cC^+_{\Delta_1,+1}\,\delta(\beta)\int_{\R^2}\frac{\d z'_1\,\d\bar z'_2\;\delta(z'_1-z_2)\,\delta(\bar z_1-\bar z'_2)}{(z'_1-z_1)^{1-\Delta_1}\,(\bar z_2'-\bar z_2)^{1-\Delta_2}}\\
        &= \frac{2\pi\,\cC^+_{\Delta_1,+1}\,\delta(\beta)}{z_{21}^{1-\Delta_1}\,\bar z_{12}^{1-\Delta_2}}\,.
    \end{split}
\end{equation}
Notice how the light transforms have symmetrically soaked up all the residual delta functions. On support of $\Delta_1+\Delta_2=2$, the conformal weights of the resulting correlator can be checked to be $(\frac{1-\Delta_1}{2},\frac{\Delta_1-1}{2})$, $(\frac{\Delta_2-1}{2},\frac{1-\Delta_2}{2})$ in gluons 1 and 2 respectively. These are the expected weights of the light transformed operators.

\paragraph{3 points.} Somewhat more non-trivially, let's consider examples of 3-point celestial amplitudes $\cA_3(1^{\eps_1}_{\Delta_1,\ell_1},2^{\eps_2}_{\Delta_2,\ell_2},3^{\eps_3}_{\Delta_3,\ell_3})$. We will work out the light transform of an MHV amplitude in the following configuration: $\ell_1=\ell_2=-\ell_3=-1$ and $\eps_1=\eps_2=-\eps_3=-1$. This corresponds to the process $1\;2\to 3$.

The corresponding celestial amplitude was computed in \cite{Pasterski:2017ylz},
\begin{equation}\label{cA3gluon}
    \cA_3(1^{-}_{\Delta_1,-1},2^{-}_{\Delta_2,-1},3^{+}_{\Delta_3,+1}) = 2\pi\,\delta(\beta)\,\Theta\biggl(\frac{z_{13}}{z_{12}}\biggr)\,\Theta\biggl(\frac{z_{32}}{z_{12}}\biggr)\,\frac{\delta(\bar z_{13})\,\delta(\bar z_{23})}{z_{12}^{-\Delta_3}\,z_{32}^{2-\Delta_1}\,z_{13}^{2-\Delta_2}}\,.
\end{equation}
Computing its light transforms, we find
\begin{equation}
    \begin{split}
        &\cL_3(1^{-}_{\Delta_1,-1},2^{-}_{\Delta_2,-1},3^{+}_{\Delta_3,+1}) = 2\pi\,\delta(\beta)\int_{\R^3}\frac{\d\bar z'_1}{(\bar z'_1-\bar z_1)^{1-\Delta_1}}\,\frac{\d\bar z'_2}{(\bar z'_2-\bar z_2)^{1-\Delta_2}}\,\frac{\d z'_3}{(z'_3-z_3)^{1-\Delta_3}}\\
        &\hspace{3cm}\times \Theta\biggl(\frac{z_1-z'_3}{z_1-z_2}\biggr)\,\Theta\biggl(\frac{z'_3-z_2}{z_1-z_2}\biggr)\,\frac{\delta(\bar z'_1-\bar z_3)\,\delta(\bar z'_2-\bar z_3)}{z_{12}^{-\Delta_3}\,(z'_3-z_2)^{2-\Delta_1}\,(z_1-z'_3)^{2-\Delta_2}}\\
        &= \frac{2\pi\,\delta(\beta)}{z_{12}^{-\Delta_3}\,\bar z_{31}^{1-\Delta_1}\,\bar z_{32}^{1-\Delta_2}}\int_{z_2}^{z_1}\frac{\sgn(z_{12})\;\d z'_3}{(z'_3-z_3)^{1-\Delta_3}\,(z'_3-z_2)^{2-\Delta_1}\,(z_1-z'_3)^{2-\Delta_2}}\,.
    \end{split}
\end{equation}
Again, the delta functions have been completely absorbed by the light ray integrals along $\bar z'_1$, $\bar z'_2$. Even the step functions have been used up to impose either $z_1>z'_3>z_2$ or $z_1<z'_3<z_2$. The factor of $\sgn(z_{12})$ simultaneously takes care of the orientation of the integral in both cases.

We have already experienced similar conformal integrals when computing light transforms of conformal primary wavefunctions in section \ref{sec:lt}. Substituting
\begin{equation}\label{3ptsub}
    z'_3 = z_2 + z_{12}\,y
\end{equation}
reduces this to a hypergeometric integral,
\begin{equation}\label{hypint}
    \begin{split}
        &\frac{2\pi\,\delta(\beta)\,\sgn(z_{12})}{z_{23}^{1-\Delta_3}\,\bar z_{31}^{1-\Delta_1}\,\bar z_{32}^{1-\Delta_2}}\int_0^1\d y\;y^{\Delta_1-2}\,(1-y)^{\Delta_2-2}\,\biggl(1-\frac{z_{21}}{z_{23}}\,y\biggr)^{\Delta_3-1}\\
        &= \frac{2\pi\,\delta(\beta)\,\sgn(z_{12})}{z_{23}^{1-\Delta_3}\,\bar z_{31}^{1-\Delta_1}\,\bar z_{32}^{1-\Delta_2}}\,B(\Delta_1-1,\Delta_2-1)\,{}_2F_1\biggl(\begin{matrix}\Delta_1-1\;,\;1-\Delta_3\\\Delta_1+\Delta_2-2\end{matrix}\,\biggl|\,\frac{z_{21}}{z_{23}}\biggr)\,.
    \end{split}
\end{equation}
It turns out that since $\Delta_1+\Delta_2-2=1-\Delta_3$ on the support of $\delta(\beta)$, the ${}_2F_1$ hypergeometric function degenerates to give a much simpler, conformally covariant final result
\begin{equation}\label{L3gluon}
    \cL_3(1^{-}_{\Delta_1,-1},2^{-}_{\Delta_2,-1},3^{+}_{\Delta_3,+1}) = \frac{2\pi\,\delta(\beta)\,B(\Delta_1-1,\Delta_2-1)\,\sgn(z_{12})}{z_{13}^{\Delta_1-1}\,z_{23}^{\Delta_2-1}\,\bar z_{31}^{1-\Delta_1}\,\bar z_{32}^{1-\Delta_2}}\,.
\end{equation}
With this, we have finally brought the 3-gluon celestial amplitude to the standard form of a 3-point CFT correlator. The conformal weights are again straightforwardly verified to be $(\frac{\Delta_1-1}{2},\frac{1-\Delta_1}{2})$, $(\frac{\Delta_2-1}{2},\frac{1-\Delta_2}{2})$, $(\frac{1-\Delta_3}{2},\frac{\Delta_3-1}{2})$ in gluons 1, 2, 3 respectively. The absence of $z_{12}$ and $\bar z_{12}$ from the denominator might seem odd at first, but they can be naturally restored by multiplying with appropriate factors of $z_{12}^\beta=\bar z_{12}^\beta=1$.

We can also consider $\overline{\text{MHV}}$ celestial amplitudes; as an example, take
\begin{equation}\label{cA3gluon1}
    \cA_3(1^{+}_{\Delta_1,+1},2^{+}_{\Delta_2,+1},3^{-}_{\Delta_3,-1}) = 2\pi\,\delta(\beta)\,\Theta\biggl(\frac{\bar z_{13}}{\bar z_{12}}\biggr)\,\Theta\biggl(\frac{\bar z_{32}}{\bar z_{12}}\biggr)\,\frac{\delta(z_{13})\,\delta(z_{23})}{\bar z_{12}^{-\Delta_3}\,\bar z_{32}^{2-\Delta_1}\,\bar z_{13}^{2-\Delta_2}}\,.
\end{equation}
Noting that our prescription of ambidextrously light transforming the celestial amplitude was parity symmetric, we can easily convince ourselves that the light transform of \eqref{cA3gluon1} is the parity conjugate of \eqref{L3gluon}
\begin{equation}\label{L3gluon1}
    \cL_3(1^{+}_{\Delta_1,+1},2^{+}_{\Delta_2,+1},3^{-}_{\Delta_3,-1}) = \frac{2\pi\,\delta(\beta)\,B(\Delta_1-1,\Delta_2-1)\,\sgn(\bar z_{12})}{z_{31}^{1-\Delta_1}\,z_{32}^{1-\Delta_2}\,\bar z_{13}^{\Delta_1-1}\,\bar z_{23}^{\Delta_2-1}}\,.
\end{equation}
It has weights $(\frac{1-\Delta_1}{2},\frac{\Delta_1-1}{2})$, $(\frac{1-\Delta_2}{2},\frac{\Delta_2-1}{2})$, $(\frac{\Delta_3-1}{2},\frac{1-\Delta_3}{2})$.

Analogous calculations for other configurations of incoming and outgoing gluons work similarly. For example, all MHV 3-point celestial amplitudes $\cA_3(1^{\eps_1}_{\Delta_1,-1},2^{\eps_2}_{\Delta_2,-1},3^{\eps_3}_{\Delta_3,+1})$ contain the same set of delta functions $\delta(\bar z_{13})\,\delta(\bar z_{23})$ \cite{Pasterski:2017ylz}. These get uniformly absorbed by our prescription of light transforms. In fact, we demonstrate this phenomenon more generally by deriving Grassmannian formulae for light transformed celestial amplitudes in section \ref{sec:grass} that are completely free of any leftover delta functions. But unlike the easy hypergeometric integral in \eqref{hypint} that led to \eqref{L3gluon}, explicit evaluation of the leftover light transform integrals requires more intricate prescriptions for choosing branches of integrands. We leave their detailed study to the future.


\subsection{Examples of graviton amplitudes}

The analysis for 2- and 3-graviton celestial amplitudes can also be performed along the same lines and is instructive to see universal features of light transformed amplitudes.

\paragraph{2 points.} Starting with the inner product of two graviton boost eigenstates (which is structurally identical to that of gluons),
\begin{equation}\label{cA2grav}
    \cA_2(1_{\Delta_1,+2}^{+},2_{\Delta_2,-2}^{-}) = 2\pi\,\cC^+_{\Delta_1,+2}\,\delta(\beta)\,\delta(z_{12})\,\delta(\bar z_{12})\,,
\end{equation}
we discover the symmetrically light transformed amplitude
\begin{equation}\label{L2grav}
    \begin{split}
        \cL_2(1_{\Delta_1,+2}^{+},2_{\Delta_2,-2}^{-}) &= 2\pi\,\cC^+_{\Delta_1,+2}\,\delta(\beta)\int_{\R^2}\frac{\d z'_1\,\d\bar z'_2\;\delta(z'_1-z_2)\,\delta(\bar z_1-\bar z'_2)}{(z'_1-z_1)^{-\Delta_1}\,(\bar z_2'-\bar z_2)^{-\Delta_2}}\\
        &= \frac{2\pi\,\cC^+_{\Delta_1,+2}\,\delta(\beta)}{z_{21}^{-\Delta_1}\,\bar z_{12}^{-\Delta_2}}\,.
    \end{split}
\end{equation}
On the support of $\Delta_1+\Delta_2=2$, this has weights $(-\frac{\Delta_1}{2},\frac{\Delta_1-2}{2})$ in the first graviton and $(\frac{\Delta_2-2}{2},-\frac{\Delta_2}{2})$ in the second.

\paragraph{3 points.} 3-graviton celestial amplitudes were derived in \cite{Puhm:2019zbl}. For simplicity, consider the same helicity configuration that we studied for gluons,
\begin{equation}\label{cA3grav}
    \cA_3(1^{-}_{\Delta_1,-2},2^{-}_{\Delta_2,-2},3^{+}_{\Delta_3,+2}) = 2\pi\,\delta(\beta+\im)\,\Theta\biggl(\frac{z_{13}}{z_{12}}\biggr)\,\Theta\biggl(\frac{z_{32}}{z_{12}}\biggr)\,\frac{\delta(\bar z_{13})\,\delta(\bar z_{23})}{z_{12}^{-2-\Delta_3}\,z_{32}^{2-\Delta_1}\,z_{13}^{2-\Delta_2}}\,.
\end{equation}
Delta functions like $\delta(\beta+\im)$ were given a practical definition in \cite{Donnay:2020guq}. Here, it essentially imposes the constraint $\Delta_1+\Delta_2+\Delta_3=2$. 

The corresponding light transform reads
\begin{align}
    &\cL_3(1^{-}_{\Delta_1,-2},2^{-}_{\Delta_2,-2},3^{+}_{\Delta_3,+2}) = 2\pi\,\delta(\beta+\im)\int_{\R^3}\frac{\d\bar z'_1}{(\bar z'_1-\bar z_1)^{-\Delta_1}}\,\frac{\d\bar z'_2}{(\bar z'_2-\bar z_2)^{-\Delta_2}}\,\frac{\d z'_3}{(z'_3-z_3)^{-\Delta_3}}\nonumber\\
    &\hspace{2.5cm}\times \Theta\biggl(\frac{z_1-z'_3}{z_1-z_2}\biggr)\,\Theta\biggl(\frac{z'_3-z_2}{z_1-z_2}\biggr)\,\frac{\delta(\bar z'_1-\bar z_3)\,\delta(\bar z'_2-\bar z_3)}{z_{12}^{-2-\Delta_3}\,(z'_3-z_2)^{2-\Delta_1}\,(z_1-z'_3)^{2-\Delta_2}}\\
    &= \frac{2\pi\,\delta(\beta+\im)}{z_{12}^{-2-\Delta_3}\,\bar z_{31}^{-\Delta_1}\,\bar z_{32}^{-\Delta_2}}\int_{z_2}^{z_1}\frac{\sgn(z_{12})\;\d z'_3}{(z'_3-z_3)^{-\Delta_3}\,(z'_3-z_2)^{2-\Delta_1}\,(z_1-z'_3)^{2-\Delta_2}}\,.\nonumber
\end{align}
Repeating the substitution \eqref{3ptsub}, one arrives at the answer
\begin{equation}\label{L3grav}
    \cL_3(1^{-}_{\Delta_1,-2},2^{-}_{\Delta_2,-2},3^{+}_{\Delta_3,+2}) = \frac{2\pi\,\delta(\beta+\im)\,B(\Delta_1-1,\Delta_2-1)\,|z_{12}|}{z_{13}^{\Delta_1-1}\,z_{23}^{\Delta_2-1}\,\bar z_{31}^{-\Delta_1}\,\bar z_{32}^{-\Delta_2}}\,,
\end{equation}
where $|z_{12}|$ is the absolute value of $z_{12}$. This has weights $(\frac{\Delta_1-2}{2},-\frac{\Delta_1}{2})$, $(\frac{\Delta_2-2}{2},-\frac{\Delta_2}{2})$ and $(-\frac{\Delta_3}{2},\frac{\Delta_3-2}{2})$ in the first, second and third graviton respectively. 

We can similarly light transform the $\overline{\text{MHV}}$ 3-graviton amplitude
\begin{equation}\label{cA3grav1}
    \cA_3(1^{+}_{\Delta_1,+2},2^{+}_{\Delta_2,+2},3^{-}_{\Delta_3,-2}) = 2\pi\,\delta(\beta+\im)\,\Theta\biggl(\frac{\bar z_{13}}{\bar z_{12}}\biggr)\,\Theta\biggl(\frac{\bar z_{32}}{\bar z_{12}}\biggr)\,\frac{\delta(z_{13})\,\delta(z_{23})}{\bar z_{12}^{-2-\Delta_3}\,\bar z_{32}^{2-\Delta_1}\,\bar z_{13}^{2-\Delta_2}}\,,
\end{equation}
finding the conjugate result
\begin{equation}\label{L3grav1}
    \cL_3(1^{+}_{\Delta_1,+2},2^{+}_{\Delta_2,+2},3^{-}_{\Delta_3,-2}) = \frac{2\pi\,\delta(\beta+\im)\,B(\Delta_1-1,\Delta_2-1)\,|\bar z_{12}|}{z_{31}^{-\Delta_1}\,z_{32}^{-\Delta_2}\,\bar z_{13}^{\Delta_1-1}\,\bar z_{23}^{\Delta_2-1}}\,.
\end{equation}
Its weights are read off to be $(-\frac{\Delta_1}{2},\frac{\Delta_1-2}{2})$, $(-\frac{\Delta_2}{2},\frac{\Delta_2-2}{2})$ and $(\frac{\Delta_3-2}{2},-\frac{\Delta_3}{2})$ in gravitons 1, 2, 3 respectively. Hence, once again we have gotten rid of all the kinematic delta functions that were present to impose residual momentum conservation. Notice also that the only substantial change in going from 3-gluon to 3-graviton light transformed amplitudes is the replacement of the factors $\sgn(z_{12})$, $\sgn(\bar z_{12})$ by the absolute values $|z_{12}|$, $|\bar z_{12}|$. This is reminiscent of the difference between the 3-gluon and 3-graviton twistor amplitudes given in equations (3.2), (3.12) of \cite{ArkaniHamed:2009si}.

\subsection{Examples of light transform OPE}
\label{sec:ltope}

Celestial amplitudes are conjectured to be the correlators of a 2d celestial CFT,
\begin{equation}
    \cA_n(1^{\eps_1}_{\Delta_1,\ell_1},2^{\eps_2}_{\Delta_2,\ell_2},\dots,n^{\eps_n}_{\Delta_n,\ell_n}) = \left\la\prod_{i=1}^n\cO_{h_i,\bar h_i}^{\eps_i}(z_i,\bar z_i)\right\ra\,.
\end{equation}
Any generic CFT can be described by the OPE algebra of its operator spectrum, and one expects this viewpoint to also apply to a CCFT. Thus, the knowledge of a complete celestial OPE algebra of the form
\begin{equation}
    \cO_{h_i,\bar h_i}^{\eps_i}(z_i,\bar z_i)\,\cO_{h_j,\bar h_j}^{\eps_j}(z_j,\bar z_j)\sim\sum_I C_{ijI}(z_{ij},\bar z_{ij},\p_j,\bar\p_j)\,\cO_I(z_j,\bar z_j)
\end{equation}
could entail an entirely holographic encoding of all of perturbative physics. 

Though this dream is far from being fully realized, recently there has been some concrete progress in determining such celestial OPE to leading and subleading orders \cite{Fan:2019emx, Pate:2019lpp, Banerjee:2020kaa, Banerjee:2020zlg, Ebert:2020nqf, Banerjee:2020vnt, Guevara:2021abz}. The main idea is that the OPE limits $z_i\to z_j$ or $\bar z_i\to\bar z_j$ coincide with the collinear limits of momentum eigenstates. Commuting these limits past the Mellin integrals in \eqref{celamp}, one is led to the much simpler job of Mellin transforming the collinear expansion of momentum space amplitudes. For instance, denoting the CCFT operators dual to outgoing gluons by $O^{a}_{\Delta,\ell=\pm}$, the celestial OPE of two positive helicity gluons is found to be \cite{Fan:2019emx, Pate:2019lpp}
\begin{equation}\label{celopgluon}
    O^a_{\Delta_1,+}(z_1,\bar z_1)\,O^b_{\Delta_2,+}(z_2,\bar z_2) \sim \frac{f^{abc}}{z_{12}}\,B(\Delta_1-1,\Delta_2-1)\,O^c_{\Delta_1+\Delta_2-1,+}(z_2,\bar z_2)+\cdots\,,
\end{equation}
where $a,b$, etc.\ are gluon color indices and $f^{abc}$ are the structure constants of the gauge Lie algebra. The leading primary on the right is again an outgoing, positive helicity gluon but with weight $\Delta_1+\Delta_2-1$. Similarly, letting $G_{\Delta,\ell=\pm}$ denote CCFT operators dual to outgoing graviton states, one finds the positive helicity OPE \cite{Pate:2019lpp}
\begin{equation}\label{celopgrav}
    G_{\Delta_1,+}(z_1,\bar z_1)\,G_{\Delta_2,+}(z_2,\bar z_2) \sim \frac{\bar z_{12}}{z_{12}}\,B(\Delta_1-1,\Delta_2-1)\,G_{\Delta_1+\Delta_2,+}(z_2,\bar z_2)+\cdots\,.
\end{equation}
The authors of \cite{Pate:2019lpp} also showed that the displayed leading OPEs are uniquely fixed by asymptotic symmetries, bolstering the holographic interpretation of collinear singularities.

Now, if we believe that \eqref{celopgluon} provides the OPE of a consistent CFT dual to gauge theory in the bulk, then we can use it expand 3-point functions in terms of 2-point functions. Writing this in terms of color-stripped gluon celestial amplitudes, one requires
\begin{equation}\label{cA3expgluon}
    \cA_3(1^+_{\Delta_1,+1},2^+_{\Delta_2,+1},3^-_{\Delta_3,-1}) \sim \frac{1}{z_{12}}\,B(\Delta_1-1,\Delta_2-1)\,\cA_2(2^+_{\Delta_1+\Delta_2-1,+1}, 3^-_{\Delta_3,-1}) + \cdots\,.
\end{equation}
And a similar expansion should also follow for gravity from the OPE \eqref{celopgrav},
\begin{equation}\label{cA3expgrav}
    \cA_3(1^+_{\Delta_1,+2},2^+_{\Delta_2,+2},3^-_{\Delta_3,-2}) \sim \frac{\bar z_{12}}{z_{12}}\,B(\Delta_1-1,\Delta_2-1)\,\cA_2(2^+_{\Delta_1+\Delta_2,+2}, 3^-_{\Delta_3,-2}) + \cdots\,.
\end{equation}
But these asymptotics are nowhere near being obvious from the highly distributional expressions \eqref{cA2gluon}, \eqref{cA3gluon1}, \eqref{cA2grav} and \eqref{cA3grav1} for these celestial amplitudes. 
Perhaps the most vexing puzzle is: \emph{where is the Beta function $B(\Delta_1-1,\Delta_2-1)$ hiding?}

\medskip

In the absence of a clear answer to these questions, we could alternatively ask if the celestial OPE of our light transformed primaries may be a better thing to look at. The main motivation for this suggestion comes from the fact that, as we saw in the previous subsections, the low multiplicity light transformed celestial amplitudes were no longer distributional in celestial kinematics. In fact, using \eqref{L2gluon} and \eqref{L3gluon1}, one can concretely demonstrate the small $z_{12}$, $\bar z_{12}$ asymptotics
\begin{multline}
    \cL_3(1^+_{\Delta_1,+1},2^+_{\Delta_2,+1},3^-_{\Delta_3,-1}) \sim  \frac{2\pi\,\delta(\beta)\,B(\Delta_1-1,\Delta_2-1)\,\sgn(\bar z_{12})}{z_{32}^{2-\Delta_1-\Delta_2}\,\bar z_{23}^{1-\Delta_3}} + \cdots\\
    \sim \sgn(\bar z_{12})\;\frac{B(\Delta_1-1,\Delta_2-1)}{\cC^+_{\Delta_1+\Delta_2-1,+1}}\,\cL_2(2^+_{\Delta_1+\Delta_2-1,+1}, 3^-_{\Delta_3,-1}) + \cdots\,.
\end{multline}
The corresponding result for gravitons follows from \eqref{L2grav} and \eqref{L3grav1}:
\begin{multline}
    \cL_3(1^+_{\Delta_1,+2},2^+_{\Delta_2,+2},3^-_{\Delta_3,-2}) \sim  \frac{2\pi\,\delta(\beta+\im)\,B(\Delta_1-1,\Delta_2-1)\,|\bar z_{12}|}{z_{32}^{-\Delta_1-\Delta_2}\,\bar z_{23}^{-\Delta_3}} + \cdots\\
    \sim |\bar z_{12}|\;\frac{B(\Delta_1-1,\Delta_2-1)}{\cC^+_{\Delta_1+\Delta_2,+2}}\,\cL_2(2^+_{\Delta_1+\Delta_2,+2}, 3^-_{\Delta_3,-2}) + \cdots\,.
\end{multline}
From these expansions, we can neatly extract the leading order light transform OPE without appealing to any collinear properties of momentum space amplitudes. We get
\begin{multline}\label{gluonltope}
    \bL[O^a_{\Delta_1,+}](z_1,\bar z_1)\,\bL[O^b_{\Delta_2,+}](z_2,\bar z_2)\\
    \sim \sgn(\bar z_{12})\;\frac{B(\Delta_1-1,\Delta_2-1)}{\cC^+_{\Delta_1+\Delta_2-1,+1}}\,f^{abc}\,\bL[O^c_{\Delta_1+\Delta_2-1,+}](z_2,\bar z_2) + \cdots
\end{multline}
for two outgoing positive helicity gluons, along with
\begin{multline}\label{gravltope}
    \bL[G_{\Delta_1,+}](z_1,\bar z_1)\,\bL[G_{\Delta_2,+}](z_2,\bar z_2)\\
    \sim |\bar z_{12}|\;\frac{B(\Delta_1-1,\Delta_2-1)}{\cC^+_{\Delta_1+\Delta_2,+2}}\,\bL[G_{\Delta_1+\Delta_2,+}](z_2,\bar z_2) + \cdots\,.
\end{multline}
for outgoing positive helicity gravitons.\footnote{The freedom in $\cC^+_{\Delta_1+\Delta_2-1,+1}$ and $\cC^+_{\Delta_1+\Delta_2,+2}$ is one of normalization of the OPE. Since they only depend on the weights in the combination $\Delta_1 +\Delta_2$, they cannot be used to completely cancel the Beta function $B(\Delta_1-1,\Delta_2-1)$ in the OPE coefficient.} Notice how these OPE are actually \emph{regular} in the OPE limit! The singularity structure has been reduced to a simple --- if somewhat mysterious from a CFT viewpoint --- sign-discontinuity. 

Entirely analogous OPE hold for the $\bar\bL$ transforms of negative helicity gluons and gravitons. And one should also be able to extend such 3-point computations to determine the light transform OPE in all other helicity and directional configurations. Of course, this is not a proof that this OPE will hold in higher multiplicity amplitudes in a universal fashion. However, in trying to find a general proof, we land on new computational hurdles. Instead of the 3-point functions, we can try directly light transforming the OPE of conformal primaries. But unlike the Mellin transform, the light transform involves integration over celestial coordinates. To be clear, consider a correlation function involving light transforms of two primaries $\cO_1$, $\cO_2$,
\begin{multline}
    \left\la\bL[\cO_1](z_1,\bar z_1)\,\bL[\cO_2](z_2,\bar z_2)\,\Phi\right\ra \\
    = \int_{\R^2}\frac{\d z_1'}{(z_1'-z_1)^{2-2h_1}}\,\frac{\d z_2'}{(z_2'-z_2)^{2-2h_2}}\;\left\la\cO_1(z_1',\bar z_1)\,\cO_2(z_2',\bar z_2)\,\Phi\right\ra\,,
\end{multline}
with $\Phi$ being a generic composite operator. Even if we take a $\bL[\cO_1]\,\bL[\cO_2]$ OPE limit $z_1\to z_2$ and naively commute it with the light ray integrals, this does not extract the singular behavior of $\la\cO_1(z_1',\bar z_1)\,\cO_2(z_2',\bar z_2)\,\Phi\ra$ as $z_1'\to z_2'$ in any simple way. Due to this, the light transform OPE depends non-locally on the conformal primary OPE. We may not be able to extract it by light transforming just a truncated set of terms in the $\cO_1\,\cO_2$ OPE.\footnote{This issue does not arise when computing the OPE of shadow transforms of symmetry currents because there the OPE expansion is replaced by exact Ward identities.\cite{Fotopoulos:2019vac, Fan:2020xjj,Fotopoulos:2020bqj}.}

Nonetheless, there are clear indications that our OPE may hold more generally. If for argument's sake we allow ourselves to naively light transform the OPE \eqref{celopgluon} term by term, then it is reassuring that the leading operator in \eqref{gluonltope} is precisely $\bL[O^c_{\Delta_1+\Delta_2-1}]$: the light transform of the leading operator $O^c_{\Delta_1+\Delta_2-1}$ on the right in \eqref{celopgluon}. An even more surprising plot twist of our computation has been that precisely the same Beta function $B(\Delta_1-1,\Delta_2-1)$ that was present in \eqref{celopgluon} has been generated by the light transforms and is present in \eqref{gluonltope}. This is also consistent with a naive light transform of the right hand side of \eqref{celopgluon} (modulo the normalization $\cC_{\Delta,\ell}^\eps$ that needs to be determined). And similar phenomena are also displayed by the graviton OPE above. The only fact that remains obscure is the procedure by which the singularities $1/z_{12}$, $\bar z_{12}/z_{12}$ get replaced by $\sgn(\bar z_{12})$, $|\bar z_{12}|$ in the two OPEs. We leave these exciting directions of investigations to future work.


\section{Grassmannian formulae}
\label{sec:grass}

In this section we will content ourselves with outlining just the basic derivation of Grassmannian representations for celestial and light transformed gluon amplitudes, our main purpose being to highlight some of the structure of the latter. As in the discovery of the original Grassmannian formula in \cite{ArkaniHamed:2009dn}, this is the natural next step in our experimentation with ambidextrous bases of light transforms. We will show that one can beautifully localize all the Mellin integrals in the former case, as well as the light ray integrals in the latter, against the Grassmannian delta functions. Such a Grassmannian formulation can open the window into a new positive geometric understanding of celestial amplitudes. We plan to return to this point in greater detail in the future \cite{Parisi:2021tba}, studying examples and generalizations of our formulae in maximally supersymmetric settings.

\subsection{For momentum space amplitudes}

The superamplitudes of $\cN=4$ SYM possess remarkable Grassmannian representations that are manifestly built out of ``on-shell processes'' \cite{Arkani-Hamed:2016byb}. The Grassmannian $\Gr(k,n)$ is the $k(n-k)$-dimensional space of maximal rank $k\times n$ complex matrices $C$ taken modulo $\GL(k)$ action.\footnote{One usually works with complex matrices $C$ irrespective of space-time signature.} Concretely, the formulae take the form of certain integrals over $\Gr(k,n)$ along contours prescribed by BCFW recursion relations. Tree-level color-ordered $n$-gluon amplitudes (with $n\geq3$) can be extracted with a particular gauge fixing of $C$. 

In what follows, we will index positive helicity particles by $\sa,\msf{b},\dots$, and negative helicity particles by $\bsa,\bar{\msf{b}},\dots$; whereas a generic particle will continue to be indexed by $i,j$, etc. We use the $\GL(k)$ freedom of the matrix $C\equiv (C_{\bsa i})\in\Gr(k,n)$ to choose
\begin{equation}
    C_{\bsa\bar{\msf{b}}} = \delta_{\bsa\bar{\msf{b}}}\,,\qquad C_{\bsa\sa} = c_{\bsa\sa}\,.
\end{equation}
That is, we have gauge fixed the $k$ columns corresponding to negative helicity particles to form the $k\times k$ identity matrix. The $k(n-k)$ non-trivial entries $c_{\bsa\sa}$ are sometimes referred to as ``link variables''. The momentum space $n$-gluon amplitude for this configuration is then given by the ``link representation'' 
\begin{equation}\label{Ank}
    A_{n,k} = \int_\Gamma\Omega_{n,k}(C)\;\delta^{2k}(C\cdot\bar\lambda)\;\delta^{2(n-k)}(C^\perp\cdot\lambda)\,.
\end{equation}
In this formula, the measure over $\Gr(k,n)$ reads
\begin{equation}
    \Omega_{n,k}(C) = \frac{\d^{k(n-k)}c}{(1\,2\,\cdots\,k)\,(2\,3\,\cdots\,k+1)\cdots(n\,1\,\cdots\,k-1)}\,,
\end{equation}
with $(i\;i+1\,\cdots\,i+k-1)$ standing for a $k\times k$ minor of $C$ involving the columns labeled $i,i+1,\dots,i+k-1$. Lastly, the notation $C^\perp\equiv(C^\perp_{\sa i})$ stands for the $(n-k)\times n$-dimensional orthogonal complement matrix with components
\begin{equation}
    C^\perp_{\sa\bsa} = -c_{\bsa\sa}\,,\qquad C^\perp_{\sa\msf{b}} = \delta_{\sa\msf{b}}\,.
\end{equation}
It satisfies $C^\perp\cdot C^t=0$, whence its rows span the vector space orthogonal to that spanned by the rows of $C$. 

Explicitly writing out the gauge fixed form of $C$, we can make the formula \eqref{Ank} a lot more tangible,
\begin{equation}\label{Ank1}
    A_{n,k} = \int_\Gamma\Omega_{n,k}(C)\;\prod_{\bsa}\delta ^2\biggl(\bar\lambda_{\bsa}+\sum_{\msf{b}}c_{\bsa\msf{b}}\,\bar\lambda_{\msf{b}}\biggr)\,\prod_{\sa}\delta^2\biggl(\lambda_\sa - \sum_{\bar{\msf{b}}}c_{\bar{\msf{b}}\sa}\,\lambda_{\bar{\msf{b}}}\biggr)\,.
\end{equation}
A priori, \eqref{Ank1} has $k(n-k)$ integrals, though $2n-4$ of them are immediately localizable against the displayed delta functions.\footnote{The remaining four delta functions impose momentum conservation.} However, the contour of integration $\Gamma\subset\Gr(k,n)$ is designed so as to also localize the remaining integrals by encircling specific choices of poles in $\Omega_{n,k}(C)$ dictated by BCFW recursion/on-shell diagrammatics. 
The interested reader is encouraged to peruse the textbook treatments in \cite{Arkani-Hamed:2016byb,Elvang:2013cua}.

We are now ready to translate the formula \eqref{Ank1} to the celestial world.


\subsection{For celestial amplitudes}

To construct celestial amplitudes, we would like to use the delta functions in \eqref{Ank1} to localize the Mellin integrals. But before we do that, we need a minor trick (no pun intended). 

Recall the definition of the $n$-gluon celestial amplitude,
\begin{equation}
    \cA_n(\zeta_i,\bar\zeta_i,\Delta_i,\ell_i,\eps_i) = \int_{\R_+^n}\prod_{j=1}^n\frac{\d\omega_j}{\omega_j}\,\omega_j^{\Delta_j}\;A_n(\sqrt{\omega_i}\,\zeta_i,\eps_i\,\sqrt{\omega_i}\,\bar\zeta_i,\ell_i)\,.
\end{equation}
We can use little group scaling to move the factors of $\sqrt{\omega_i}$ and signs $\eps_i$ around. Specifically, we rewrite the momentum space amplitude on the right as
\begin{equation}
    A_n(\sqrt{\omega_i}\,\zeta_i,\eps_i\,\sqrt{\omega_i}\,\bar\zeta_i,\ell_i) = A_n(\eps_{\bsa}\,\zeta_{\bsa},\omega_{\bsa}\,\bar\zeta_{\bsa},-\,;\,\omega_\sa\,\zeta_\sa,\eps_\sa\,\bar\zeta_\sa,+)\prod_{j=1}^n\omega_j\,.
\end{equation}
To be clear, we have used the following little group scalings of the associated spin 1 states:
\begin{equation}
    \begin{split}
        |\sqrt{\omega_{\bsa}}\,\zeta_{\bsa},\eps_{\bsa}\,\sqrt{\omega_{\bsa}}\,\bar\zeta_{\bsa},-\ra &= \biggl(\frac{\sqrt{\omega_{\bsa}}}{\eps_{\bsa}}\biggr)^{-2(-1)}\,|\eps_{\bsa}\,\zeta_{\bsa},\omega_{\bsa}\,\bar\zeta_{\bsa},-\ra = \omega_{\bsa}\,|\eps_{\bsa}\,\zeta_{\bsa},\omega_{\bsa}\,\bar\zeta_{\bsa},-\ra\,,\\
        |\sqrt{\omega_\sa}\,\zeta_\sa,\eps_\sa\,\sqrt{\omega_\sa}\,\bar\zeta_\sa,+\ra &= \bigg(\frac{1}{\sqrt{\omega_\sa}}\biggr)^{-2(+1)}\,|\omega_\sa\,\zeta_\sa,\eps_\sa\,\bar\zeta_\sa,+\ra = \omega_\sa\,|\omega_\sa\,\zeta_\sa,\eps_\sa\,\bar\zeta_\sa,+\ra\,.
    \end{split}
\end{equation}
Hence, the celestial amplitude becomes
\begin{equation}
    \cA_n(\zeta_i,\bar\zeta_i,\Delta_i,\ell_i,\eps_i) = \int_{\R_+^n}\prod_{j=1}^n\d\omega_j\,\omega_j^{\Delta_j}\;A_n(\eps_{\bsa}\,\zeta_{\bsa},\omega_{\bsa}\,\bar\zeta_{\bsa},-\,;\,\omega_\sa\,\zeta_\sa,\eps_\sa\,\bar\zeta_\sa,+)\,.
\end{equation}
Into this, we substitute \eqref{Ank1} with the displayed spinor-helicity data $\bar\lambda^{\dal}_{\bsa} = \omega_{\bsa}\,\bar\zeta^{\dal}_{\bsa}$, etc.

The result is
\begin{multline}
    \cA_{n,k}\equiv\cA_n(\zeta_i,\bar\zeta_i,\Delta_i,\ell_i,\eps_i) = \int_\Gamma\Omega_{n,k}(C)\int_{\R_+^n}\prod_{j=1}^n\d\omega_j\,\omega_j^{\Delta_j}\\
    \times\prod_{\bsa}\delta ^2\biggl(\omega_{\bsa}\,\bar\zeta_{\bsa}+\sum_{\msf{b}}\eps_{\msf{b}}\,c_{\bsa\msf{b}}\,\bar\zeta_{\msf{b}}\biggr)\,\prod_{\sa}\delta^2\biggl(\omega_\sa\,\zeta_\sa - \sum_{\bar{\msf{b}}}\eps_{\bar{\msf{b}}}\,c_{\bar{\msf{b}}\sa}\,\zeta_{\bar{\msf{b}}}\biggr)\,.
\end{multline}
Finally, localizing the $\omega_j$ integrals yields
\begin{equation}\label{cAnk}
    \cA_{n,k} = \int_\Gamma\Omega_{n,k}(C)\prod_{j=1}^n\Theta(E_j)\,E_j^{\Delta_j}\prod_{\bsa}\delta\biggl(\sum_{\msf{b}}\eps_{\msf{b}}\,c_{\bsa\msf{b}}\,[\bsa\,\msf{b}]\biggr)\prod_{\sa}\delta\biggl(\sum_{\bar{\msf{b}}}\eps_{\bar{\msf{b}}}\,c_{\bar{\msf{b}}\sa}\,\la\sa\,\bar{\msf{b}}\ra\biggr)\,,
\end{equation}
having defined the ``energies'' $E_j$ as the (necessarily positive) solutions for the $\omega_j$'s,
\begin{equation}\label{Evals}
    E_{\bsa} = \omega_{\bsa} = -\sum_{\msf{b}}\eps_{\msf{b}}\,c_{\bsa\msf{b}}\,\frac{[\msf{b}\,\bar\iota]}{[\bsa\,\bar\iota]}\,,\qquad E_\sa = \omega_\sa = \sum_{\bar{\msf{b}}}\eps_{\bar{\msf{b}}}\,c_{\bar{\msf{b}}\sa}\,\frac{\la\bar{\msf{b}}\,\iota\ra}{\la\sa\,\iota\ra}\,.
\end{equation}
As usual, these have been written in terms of two auxiliary reference spinors $\iota_\al$, $\bar\iota_{\dal}$ but are actually independent of them on support of the remaining delta functions. Also, we have imposed their positivity using Heaviside step functions $\Theta(E_j)$. These constrain the contour of integration and are physically interpreted as telling us the allowed channels in which the gluons may scatter.

The formulae \eqref{cAnk} and \eqref{Evals} can be further adapted to affine coordinates $z_i,\bar z_i$. The choices $\iota_\al = \bar\iota_{\dal} = (1,0)$, $\zeta_{i\,\al}=(z_i,1)$, $\bar\zeta_{i\,\dal} = (\bar z_i,1)$ reduce the various spinor contractions to 1: $[i\,\bar\iota]=1=\la i\,\iota\ra$. This leads to the affine versions of the Grassmannian formulae for celestial gluon amplitudes,
\begin{equation}\label{cAnk1}
    \cA_{n,k} = \int_\Gamma\Omega_{n,k}(C)\prod_{j=1}^n\Theta(E_j)\,E_j^{\Delta_j}\prod_{\bsa}\delta\biggl(\sum_{\msf{b}}\eps_{\msf{b}}\,c_{\bsa\msf{b}}\,\bar z_{\bsa\msf{b}}\biggr)\prod_{\sa}\delta\biggl(\sum_{\bar{\msf{b}}}\eps_{\bar{\msf{b}}}\,c_{\bar{\msf{b}}\sa}\,z_{\sa\bar{\msf{b}}}\biggr)\,,
\end{equation}
with the energies simplifying to
\begin{equation}\label{Evals1}
    E_{\bsa} = \omega_{\bsa} = -\sum_{\msf{b}}\eps_{\msf{b}}\,c_{\bsa\msf{b}}\,,\qquad E_\sa = \omega_\sa = \sum_{\bar{\msf{b}}}\eps_{\bar{\msf{b}}}\,c_{\bar{\msf{b}}\sa}\,.
\end{equation}
We remark that since we haven't performed any light transforms yet, the above formulae can potentially be analytically continued to Minkowski space by assuming Lorentzian reality conditions $\bar z_i = z^*_i$. However, one would need to carefully define the contours of integration as we no longer have the luxury to completely localize the Grassmannian integrals against the delta functions.


\subsection{For light transformed amplitudes}

The formulae \eqref{cAnk} and \eqref{cAnk1} interpolate between integrals of Euler-type and integrals that simply localize on delta functions. The resulting celestial amplitudes take the form of generalized hypergeometric functions satisfying many physically and holographically interesting differential equations \cite{Schreiber:2017jsr, Banerjee:2020vnt, Banerjee:2020zlg, Hu:2021lrx}. 
Similarly, the singularity structure of Euler-type integrals has also entered physics in the recent studies on stringy canonical forms \cite{Arkani-Hamed:2019mrd}. It is thus curious to note that, as we show below, light transforms provide the most natural translation of \eqref{cAnk} to integrals of purely Euler-type. Owing to this structure, they may be a better starting point to characterize celestial correlators purely through the differential equations they satisfy.

The light transforms in \eqref{Ldef} are most easily performed in affine coordinates $z_i,\bar z_i$ as the energies $E_j$ are then independent of the celestial positions. They are given by the simplified expressions \eqref{Evals1}. We are thus led to compute
\begin{multline}
    \cL_{n,k} = \int_\Gamma\Omega_{n,k}(C)\prod_{j=1}^n\Theta(E_j)\,E_j^{\Delta_j}\int_{\R^k}\prod_{\bsa}\frac{\d\bar z'_{\bsa}}{(\bar z'_{\bsa}-\bar z_{\bsa})^{1-\Delta_{\bsa}}}\,\delta\biggl(\sum_{\msf{b}}\eps_{\msf{b}}\,c_{\bsa\msf{b}}\,(\bar z'_{\bsa}-\bar z_{\msf{b}})\biggr)\\
    \times\int_{\R^{n-k}}\prod_\sa\frac{\d z'_\sa}{(z'_\sa-z_\sa)^{1-\Delta_\sa}}\,\delta\biggl(\sum_{\bar{\msf{b}}}\eps_{\bar{\msf{b}}}\,c_{\bar{\msf{b}}\sa}\,(z'_\sa-z_{\bar{\msf{b}}})\biggr)\,,
\end{multline}
having abbreviated $\cL_n(\zeta_i,\bar\zeta_i,\Delta_i,\ell_i,\eps_i)\equiv\cL_{n,k}$. Remembering \eqref{Evals1}, we can solve for the $\bar z'_{\bsa}$ and $z'_\sa$ by means of the remaining delta functions:
\begin{equation}\label{zvals}
    \bar z'_{\bsa} = -\frac{1}{E_{\bsa}}\sum_{\msf{b}}\eps_{\msf{b}}\,c_{\bsa\msf{b}}\,\bar z_{\msf{b}}\,,\qquad z'_\sa = \frac{1}{E_\sa}\sum_{\bar{\msf{b}}}\eps_{\bar{\msf{b}}}\,c_{\bar{\msf{b}}\sa}\,z_{\bar{\msf{b}}}\,.
\end{equation}
Consequently, localizing the light transforms against these delta functions yields
\begin{equation}\label{Lnk}
    \cL_{n,k} = \int_\Gamma\Omega_{n,k}(C)\prod_{j=1}^n\Theta(E_j)\prod_{\bsa}\biggl(\sum_{\msf{b}}\eps_{\msf{b}}\,c_{\bsa\msf{b}}\,\bar z_{\msf{b}\bsa}\biggr)^{\Delta_{\bsa}-1}\prod_{\sa}\biggl(\sum_{\bar{\msf{b}}}\eps_{\bar{\msf{b}}}\,c_{\bar{\msf{b}}\sa}\,z_{\bar{\msf{b}}\sa}\biggr)^{\Delta_\sa-1}\,.
\end{equation}
Quite beautifully, the factors of $E_j^{\Delta_j}$ have completely dropped out! This is our main result for the celestial amplitude in an ambidextrous basis of light transformed gluons.

For instance, with a little work it can be verified that this gives the same 3-point amplitudes that we studied in our examples in section \ref{sec:gluonltamp}. In that section, we computed light transforms of 2- and 3-gluon celestial amplitudes explicitly and showed that they were no longer distributional in the $z_{ij}$, $\bar z_{ij}$. Though it is obvious in hindsight, one can also see from \eqref{Lnk} that the same is true for 4-gluon amplitudes. The light transforms have soaked up all the remaining Grassmannian delta functions, including the ones that would have been present at low multiplicity to impose residual momentum conservation. In particular, this shows that the light transforms will necessarily soak up the $\delta(z-\bar z)$ singularity in the cross ratios $z=z_{12}z_{34}/z_{13}z_{24}$ that occurs in the 4-gluon celestial amplitude \cite{Pasterski:2017ylz}. It would be very interesting to study the OPE limits of such 4-gluon (also 4-graviton) light transformed amplitudes, and verify whether they are consistent with \eqref{gluonltope}, etc. It should also allow us to access subleading light transform OPE coefficients.


\section{Discussion}
\label{discussion}

This paper arose from the need to find the correct basis of ``local'' operators in which to study CCFT correlation functions. From even our basic results it is clear that this is a long road to travel, but one laden with some very interesting clues. But light transformed operators cannot be local, can they? After all, they do have non-integer 2d spins $\pm(1-\Delta)$. This is true with respect to the usual 4d Lorentz group that we are taking to be the conformal group of our CCFT. But, as initial results from \cite{Strominger:2021lvk} seem to suggest, there may be ways of twisting the conformal group or finding different copies of $\SL(2,\R)$ whose representation theories could act as better organizational principles for the CCFT spectrum. Light transformed boost eigenstates may secretly be conformal primary with respect to such alternative conformal groups. 

Our demonstration that 2- and 3-point light transformed amplitudes like \eqref{L2gluon}, \eqref{L3gluon}, etc.\ take the form of standard conformally covariant 2- and 3-point structures even when the usual celestial amplitudes \eqref{cA2gluon}, \eqref{cA3gluon}, etc.\ don't again points to this possibility. In fact, in conventional CFTs containing a lowest energy vacuum, 2- and 3-point time-ordered correlators of light transformed operators necessarily vanish. This is because light transforms of local operators annihilate the vacuum; see for example lemma 2.1 in \cite{Kravchuk:2018htv}.\footnote{The author would like to thank Murat Kologlu for pointing out this fact.} Since we found non-trivial 2- and 3-point functions, this gives more credence to the possibility that either celestial amplitudes have no interpretation as time-ordered (in a 2d sense) CCFT correlators, or that the celestial gluon and graviton operators which we started with may not be local operators after all! 


There are also a multitude of other directions for future study. Unlike the light/shadow transform OPE of conformally soft modes, finding the light/shadow transform OPE of hard conformal primaries can get much more involved. This is because it depends non-locally on the OPE of conformal primaries and would in principle need us to resum light transforms of infinitely many terms in the usual celestial OPE expansion. Nevertheless, preliminary investigations in section \ref{sec:ltope} lead us to believe that the situation may be simpler than we expect. As we showed there, the leading operator in the positive helicity light transform OPE turns out to be the light transform of the leading operator in the positive helicity conformal primary OPE. And the same Beta function OPE coefficient accompanies the leading operators in both OPEs! Hence, it appears that we may indeed be able to compute the light transform OPE by systematically light transforming the celestial gluon/graviton OPE term by term. This would be an important step in trying to understand whether light transformed boost eigenstates give the most ``fundamental'' representation of the CCFT operator algebra. In this regard, it would also be very interesting to find a conformal block decomposition of the ambidextrously light transformed 4-gluon/graviton amplitudes.

The connections between light transforms and twistor theory are also very tantalizing. Twistors play a very important role in unraveling the integrability and infinite dimensional symmetries of self-dual Yang-Mills and gravity. For instance, in twistor space, the $w_{1+\infty}$ symmetry of self-dual GR is elegantly realized as the loop group of area preserving diffeomorphisms of the 2-planes coordinatized by $\mu^{\dal}$ at given $\lambda_\al$ \cite{Boyer:1985aj, Ooguri:1991fp}. Quite surprisingly, a discrete analogue of the light transform also made an appearance in \cite{Strominger:2021lvk}. It helped in the identification of the OPE algebra of positive helicity soft theorems \cite{Guevara:2019ypd,Guevara:2021abz} with exactly the symmetries of these integrable SD subsectors. It is very plausible that this is a consequence of light transforms (essentially) mapping boost eigenstates to primary twistor eigenstates. Operators dual to conformally soft twistor eigenstates could then provide the cleanest generators of these symmetries, in the spirit of twistor and ambitwistor strings \cite{Adamo:2014yya, Adamo:2015fwa, Geyer:2014lca, Adamo:2019ipt}. We will return to the applications of twistor strings to these infinite dimensional symmetries and positive helicity soft theorems in \cite{Adamo:2021tba}.

Lastly, in section \ref{sec:grass} we have outlined the derivation of new Grassmannian formulae for tree-level gluon celestial amplitudes as well as their ambidextrous light transforms. The light transformed Grassmannian integrals take a particularly beautiful Euler-type form. Unlike having to choose one specific gluon to shadow transform as in \cite{Crawley:2021ivb,Fan:2021isc}, our procedure and formulae treat all the gluons on equal footing. Perhaps more interestingly, our Grassmannian formulae potentially connect two important branches of the field of scattering amplitudes: celestial amplitudes and positive geometries. The work in \cite{Arkani-Hamed:2020gyp} has already shown that the positive geometry of effective field theories can be naturally encoded into the analytic properties of celestial amplitudes. It is not inconceivable that various other appearances of positivity in the study of amplitudes will also leave their imprints in celestial CFTs, and vice versa. We will further explore our Grassmannian formulae in the setting of celestial superamplitudes in future studies \cite{Parisi:2021tba}. 

\acknowledgments

I would like to thank Johan Henriksson, Lionel Mason, Matteo Parisi, Sabrina Pasterski, Andrea Puhm, Ana-Maria Raclariu, Anders Schreiber, and Diandian Wang for discussions that spurred some of the questions tackled in this work. I am grateful to Tim Adamo for comments on the draft. I am supported by a Mathematical Institute Studentship, Oxford.

\bibliographystyle{JHEP}
\bibliography{ls}

\end{document}